\documentclass{article} 
\usepackage{graphicx,float}
\usepackage{t1enc}
\usepackage{color}
\usepackage{pdfsync}
\usepackage[latin1]{inputenc}
\usepackage{mathrsfs,xspace,theorem}
\usepackage{amsmath,amssymb}

\theorembodyfont{\rmfamily}

\advance\textheight by 2cm
\advance\topmargin -1cm

\advance\textwidth by 2cm
\advance\oddsidemargin by -1cm
\advance\evensidemargin by -1cm

\floatstyle{ruled}
\newfloat{Algorithm}{htpb}{alg}

\newcommand{\wellbefore}{\ll}

\newcommand{\?}{\mskip1.5mu}

\newcommand{\lst}[2]{${#1}_0$,~${#1}_1$, $\dots\,$,~${#1}_{#2-1}$}

\newcounter{prgline}
\newcommand{\pl}{\theprgline\addtocounter{prgline}{1}}

\newcounter{ifline}
\newcounter{whileendline}
\newcounter{middlewhileend}
\newcounter{whileexit}
\newcounter{innerwhileloop}
\newcounter{exitouterloop}
\newcounter{externalwhileblock}
\newcounter{internalwhileblock}
\newcounter{brouwerone}
\newcounter{brouwertwo}
\newcounter{brouwerthree}

\newenvironment{renumerate}{\begin{enumerate}}{\end{enumerate}}

\newcommand{\IF}{\text{\textbf{if}}\xspace}
\newcommand{\BEGIN}{\text{\textbf{begin}}\xspace}

\newcommand{\FOREVER}{\text{\textbf{forever}}\xspace}

\newcommand{\END}{\text{\textbf{end}}\xspace}
\newcommand{\DO}{\text{\textbf{do}}\xspace}

\newcommand{\THEN}{\text{\textbf{then}}\xspace}
\newcommand{\ELSE}{\text{\textbf{else}}\xspace}

\newcommand{\RETURN}{\text{\textbf{return}}\xspace}

\newcommand{\LOR}{\text{\textbf{or}}\xspace}
\newcommand{\LAND}{\text{\textbf{and}}\xspace}

\newcommand{\WHILE}{\text{\textbf{while}}\xspace}
\newcommand{\BREAK}{\text{\textbf{break}}\xspace}

\newcommand{\NULL}{\text{\textbf{null}}\xspace}
\newcommand{\FUNCTION}{\text{\textbf{function}}\xspace}
\newcommand{\INITIALLY}{\text{\textbf{Initially}}\xspace}
\newcommand{\PROCEDURE}{\text{\textbf{procedure}}\xspace}
\newcommand{\var}[1]{{\itshape #1}}

\newcounter{noqed}
\newcommand{\qed}{ \ifmmode\mbox{
}\fi\rule[-.05em]{.3em}{.7em}\setcounter{noqed}{0}}
\newenvironment{proof}[1][{}]{\noindent{\bf Proof#1.
}\setcounter{noqed}{1}}{\ifnum\value{noqed}=1\qed\fi\par\medskip}

\newcommand{\CCB}{Clarke--Cormack--Burkowski\xspace}

\newcommand{\AND}{\operatorname{AND}}
\newcommand{\OR}{\operatorname{OR}}
\newcommand{\BLOCK}{\operatorname{BLOCK}}
\newcommand{\ONAND}{\operatorname{AND}_{\mathalpha<}}
\newcommand{\LOWPASS}{\operatorname{LOWPASS}}

\def\..{\,\mathpunct{\ldotp\ldotp}} 

\newcommand{\ebsuff}{\trianglelefteq}

\newcommand{\sbprol}{\preceq}
\newcommand{\sbprolneq}{\prec}

\newcommand{\C}{\mathscr C}
\newcommand{\A}{\mathscr A}
\newcommand{\B}{\mathscr B}
\newcommand{\op}{{\operatorname{op}}}
\newcommand{\url}{\cite{myurl}}

\newtheorem{theorem}{Theorem}

\newtheorem{definition}{Definition}

\renewcommand{\epsilon}{\varepsilon}
\renewcommand{\phi}{\varphi}

\title{Efficient Optimally Lazy Algorithms\\ for Minimal-Interval
Semantics\thanks{A preliminary version of some of the results in this paper
appeared in~\cite{BoVELAMIS}. All algorithms have been significantly
simplified, getting rid of the double queues of~\cite{BoVELAMIS}.
The notion of optimal laziness and all related results are entirely new.
The algorithm for the AND operator has been improved during the proofs of optimality (it was not optimally lazy in the formulation of~\cite{BoVELAMIS}).}}
\author{Paolo Boldi \qquad Sebastiano Vigna\\Dipartimento di Informatica,
Universit\`a degli Studi di Milano}
\date{}

\begin{document}
\bibliographystyle{plain}
\maketitle

\begin{abstract}
Minimal-interval semantics~\cite{CCBASTSFI} associates with each query over a
document a set of intervals, called \emph{witnesses},
that are incomparable with respect to inclusion (i.e., they form an
antichain):
witnesses define the minimal regions of the document satisfying the query.
Minimal-interval semantics makes it easy to define and compute several
sophisticated proximity
operators, provides snippets for user presentation, and can be used to rank
documents. In this paper we provide algorithms for computing conjunction and
disjunction that are linear in the number of intervals and logarithmic in the number of operands;
for additional operators, such as ordered conjunction and Brouwerian difference,
we provide linear algorithms. In all cases, space is linear in the number of
operands.
More importantly, we define a formal property of \emph{optimal
laziness}, and
either prove it, or prove its impossibility, for each algorithm. We cast our
results in a general framework of finite antichains of intervals on
total orders, making our algorithms directly applicable to other domains.
\end{abstract}


\section{Introduction}

Modern information-retrieval systems, such as web search engines, rely
on \emph{query expansion}, an automatic or semi-automatic mechanism that aims at
rewriting the user intent (i.e., a set of keywords, maybe with additional
context such as geographical location, past search history, etc.) as a
structured query built upon a number of operators. The simplest case is that
of the \emph{Boolean model}, in which operators are just conjunction,
disjunction and negation: as an example, the set of keywords provided by the
user might be expanded as disjunctions of syntactically or semantically similar
terms, and finally such disjunctive queries would be connected using
a conjunction operator. The semantics provided by this model is simply either 
the value ``true'' or the value ``false'' (in the example, ``true'' is returned
if the document contains at least one term from each disjunction).

When a document satisfies a query, however, the Boolean model fails to explain
\emph{why} and \emph{where} the query is satisfied: this information is
compressed in a single truth value. 
\emph{Minimal-interval semantics} is a richer semantic model that
uses \emph{antichains\footnote{An \emph{antichain} of a partial order is a set
of elements that are pairwise incomparable.} of intervals of natural numbers} to represent the semantics of
a query; this is the natural framework in which operators such as ordered
conjunction, proximity restriction, etc., can be defined and combined freely. Each
interval is a \emph{witness} of the satisfiability of the query,
and defines a region of the document that satisfies the query (positions in the
document are numbered starting from $0$, so regions of text are identified
with sets of consecutive integers, a.k.a.~intervals).

Consider, for example, the document given by the start
of the well-known rhyme
\begin{quote}
\emph{Pease porridge hot! Pease porridge cold!}
\end{quote}
If we query this document with the keyword ``\emph{hot}'', we just get the
Boolean answer ``true''. But a more precise answer would be ``\emph{hot} appears
in the third position of the document'': formally, this is described by the
interval $[2\..2]$.
Sometimes, a query will be satisfied in multiple parts
of the document; for example, the query ``\emph{pease}'' has answer
$\{\?[0\..0],[3\..3]\?\}$. If we consider \emph{two} keywords things become more
interesting: when we submit the conjunctive query ``\emph{pease} AND
\emph{porridge}'' the answer will be
$A=\{\?[0\..1],[1\..3],[3\..4]\?\}$.
Of course, there are more intervals containing both \emph{pease} and
\emph{porridge}, but they are omitted because they contain (and thus they are
less informative than) one of the intervals in $A$.
The latter observation leads us to considering sets of intervals that are
incomparable with respect to inclusion, that is, antichains with respect to $\subseteq$.

This approach has been defined and studied to its full extent by Clarke, Cormack and
Burkowski in their seminal paper~\cite{CCBASTSFI}. They showed that antichains
have a natural lattice structure that can be used to interpret conjunctions and
disjunctions in queries. Moreover, it is possible to define several additional
operators (proximity, followed-by, and so on) directly on the antichains. The
authors have also described families of successful ranking schemes based on the
number and length of the intervals involved~\cite{ClCSSRR}.

The main feature of minimal-interval semantics is that, by its very definition,
an antichain of intervals cannot contain more than $w$ intervals, where $w$ is
the number of words in the document. Thus, it is in principle possible
to compute all minimal-interval operators in time linear in the document size.
This is not true, for instance, if we consider different interval-semantics
approaches in which \emph{all} intervals are retained and indexed (e.g., the PAT
system~\cite{GonPAT} or the \texttt{sgrep} tool~\cite{JaKNTRA}), as the overall
number of output regions is quadratic in the document size.

In this paper, we attack the problem of providing efficient lazy algorithms for
the computation of a number of operators on antichains. As
a subproblem, we can compute the proximity of a set of terms, and indeed we are partly inspired by previous work
on proximity~\cite{SaIFAkWPS,NiPPAIGF}. Our algorithms are linear in the number
of input intervals. For conjunction and disjunction, there is also a
multiplicative logarithmic factor in the number of input antichains, which
however can be shown to be essentially unavoidable in the disjunctive case.
The space used by all algorithms is linear in the number of input antichains
(in fact, we need to store just one interval per antichain), so they are a very
extreme case of stream transformation algorithms~\cite{BYKSRSAACTG,HRRCDD}.
Moreover, our algorithms satisfy some stringent formal \emph{laziness}
properties.

Note that from a practical viewpoint laziness
makes our algorithms very attractive when paired with an index
structure (see, e.g., quasi-succinct indices~\cite{VigQSI}) that provides lazy
I/O for reading positions. For example, it is possible to decide that a set of
terms appear within a certain proximity bound \emph{without} reading all
positions: if the underlying index is lazy, our algorithms limit the I/O as much
as possible. In the open-source world, 
the semantic engine M\'imir~\cite{TBRM} is based on
MG4J~\cite{BoVTREC2005}, which contains our implementation of such algorithms.

In Section~\ref{sec:sem} we briefly introduce minimal-interval semantics,
and provide some examples and motivations. The presentation
is rather algebraic, and uses standard terms from mathematics and order theory
(e.g., ``interval'' instead of ``extent'' as in~\cite{CCBASTSFI}). The resulting
structure is essentially identical to that described in the original
paper~\cite{CCBASTSFI}, but our systematic approach makes 
good use of well-known results from order
theory, making the introduction self-contained. For some mathematical
background, see, for instance,~\cite{BirLT,DPILO}.

Another advantage of our approach is that by representing abstractly
regions of text as intervals of natural numbers we can easily highlight
connections with other areas of computer science: for example, antichains of
intervals have been used for testing
distributed computations~\cite{JRJCOLMAP}. The problem of computing operators
on antichains has thus an intrinsic interest that goes beyond 
information retrieval. This is the reason why we cast all our
results in the general framework of antichains of intervals 
on arbitrary (totally) ordered sets.

Finally, we present our algorithms. First we discuss algorithms based on queues,
and then greedy algorithms.\footnote{A free implementation of all algorithms described in this paper
is available as a part of MG4J~\cite{BoVTREC2005}
(\texttt{http://mg4j.di.unimi.it/}) and LaMa4J (\texttt{http://lama4j.di.unimi.it/}).}

\section{Minimal-interval semantics}
\label{sec:sem}

Given a totally ordered set $O$, 
let us denote with $\C_O$ the set of intervals\footnote{A subset $X$ of $O$ is
an \emph{interval} if $x,y\in X$ and $x\leq z\leq y$ imply $z \in X$.} of $O$ that are
either empty or of the form $[\ell\..r]=\{\?x\in O\mid \ell\leq x\leq r\?\}$.
Our working example will always be $O=W$, where $W=\{\?0,1,\ldots,w-1\?\}$ and $w$ represents the number of words
in a document, numbered starting from $0$ (see Figure~\ref{fig:example}); elements of
$\C_W$ can be thought of as regions of text.

Intervals are ordered by containment: when we want to order them
by \emph{reverse} containment instead, we shall write $\C_O^\op$ (``op'' stands
for ``opposite'').
Given intervals $I$ and $J$, the interval \emph{spanned} by $I$ and $J$ is the
least interval containing $I$ and $J$ (in fact, their least upper bound in
$\C_O$).

The idea behind minimal-interval
semantics~\cite{CCBASTSFI} is that every interval in $\C_W$ is a \emph{witness}
that a given query is satisfied by a document made of $w$ words. \emph{Smaller
witnesses imply a better match, or more information}; in particular, if an
interval is a witness, any
interval containing it is also a witness. We also expect that \emph{more
witnesses imply more information}. Thus, when expressing the semantics of a query, we
discard non-minimal intervals, as there are intervals that provide more relevant
information. As a result, minimal-interval semantics associates with each query
an \emph{antichain} of intervals. For instance, in
Figure~\ref{fig:example} we see a short passage of text, and the antichain of
intervals corresponding to a query. Note that, for
instance, the interval $[0\..3]$ is not included because it is not minimal (e.g.,
it contains $[0\..2]$).

It is however more convenient to start from an algebraic viewpoint. A
\emph{lower set} $X$ 
is a subset of a partial order that is closed downwards: if $y\leq x$ and $x\in X$, then 
$y\in X$. Given a subset $A$ of a partial order, we denote with
$\mathord\downarrow A$ the smallest lower set containing $A$ (i.e., the set
containing $A$ and all elements smaller than an element of $A$).

We are interested in computing operators of the 
distributive lattice $\A_O$ whose
elements are finite antichains of $\C_O^\op$ endowed with the following order:\footnote{$\A_O$ is a distributive lattice because
$\C_O^\op$ has binary greatest lower bounds~\cite{BoVLAFI2}.}
\[
A\leq B \iff \mathord\downarrow A\subseteq\mathord\downarrow B.
\]

The lattice of finite antichains\footnote{We remark that the construction of 
finite antichains (which are equivalent to finitely generated lower sets) of
compact elements is the first step in the concrete construction of the \emph{Hoare powerdomain}~\cite{AbJDT}.
Thus, several formulas appearing in the rest of the paper will look
familiar to readers acquainted with domain theory.}
$\A_W$ thus defined is essentially the classic
\CCB \emph{region algebra}, with the difference that since we allow
the empty interval, we have a top element that contains only the empty
interval, and which makes $\A_O$ a bounded lattice in the infinite
case, too. For the purposes of this paper, the difference is immaterial, though.

To make the reader grasp more easily the meaning of $\A_O$, we now describe
in an elementary way its order and its lattice operations, which have been
characterized in~\cite{BoVLAFI2}.
Given antichains $A$ and $B$, we have
\[
  A \leq B \iff \forall I \in A \quad \exists J \in B \quad J \subseteq I.
\]
Intuitively, $A\leq B$ if every witness $I$ in $A$ (an interval) can be
substituted by a better (or equal) witness $J$ in $B$, where ``better'' means
that the new witness $J$ is contained in $I$.

Correspondingly, the $\lor$ of two antichains $A$ and $B$ is given by the union
of the intervals in $A$ and $B$ from which non-minimal intervals have been
eliminated. Finally, the $\land$ of $A$ and $B$ is given by the set of all
intervals spanned by pairs of intervals $I\in A$ and $J\in B$, from which
non-minimal intervals have been eliminated. It is this very natural algebraic
structure that has led to the definition of the \CCB lattice.

We remark that the intervals in an antichain can be ordered in
principle either by left or by right extreme, but these orders can be easily
shown to be the same, so we can say that the intervals in an antichain are
naturally linearly ordered by their extremes.

\subsection{Examples}

\begin{figure}
\begin{center}
\includegraphics[scale=.7]{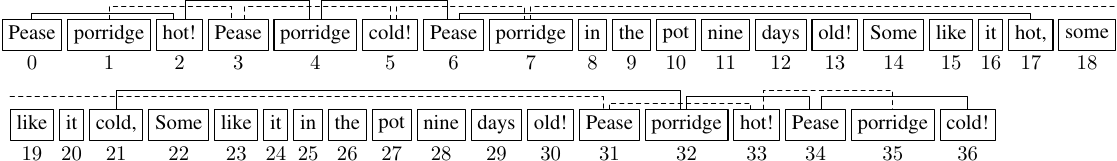}
\end{center}
\caption{\label{fig:example}A sample text; the intervals corresponding to the
semantics of the query ``\emph{porridge}
AND
\emph{pease} AND (\emph{hot} OR \emph{cold})'' are shown. For easier
reading, every other interval is dashed.}
\end{figure}

Consider from Figure~\ref{fig:example} the positions of ``\emph{porridge}''
$(1,4,7,32,35)$, ``\emph{pease}'' $(0,3,6,31,34)$, ``\emph{hot}'' $(2,17,33)$ and  ``\emph{cold}''
$(5,21,36)$. Queries associated with a single keyword have an easy
semantics---the list of positions as singleton intervals. For example, the
semantics of ``\emph{hot}'' will be 
\[
\{\?[2\..2],[17\..17],[33\..33]\?\}.
\]
If we start combining terms disjunctively, we get simply the union of their
positions.
For instance, ``\emph{hot} OR \emph{cold}'' gives
\[
\{\?[2\..2],[5\..5],[17\..17],[21\..21],[33\..33],[36\..36]\?\}.
\]
If we consider the conjunction of two terms, we will start getting non-singleton
intervals: the semantics of ``\emph{pease} AND \emph{porridge}'' is computed by
picking all possible pairs of positions of \emph{pease} and \emph{porridge}
and keeping the minimal intervals among those spanned by such pairs:
\[
\{\?[0\..1],[1\..3],[3\..4],[4\..6],[6\..7],[7\..31],[31\..32],[32\..34],[34\..35]\?\}.
\]
The more complex query ``(\emph{pease} AND \emph{porridge}) OR \emph{hot}'' is
interesting because we have to take the intervals just computed, put them
together with the positions of \emph{hot}, and remove the non-minimal intervals:
\[
\{\?[0\..1],[2\..2],[3\..4],[4\..6],[6\..7],[17\..17],[31\..32],[33\..33],[34\..35]\?\}.
\]
One can see, for example, that the presence of \emph{hot} in position $2$ has
eliminated the interval $[1\..3]$.

Let's try something more complex: ``\emph{pease} AND \emph{porridge} AND (\emph{hot} OR \emph{cold})''.
We have again to pick one interval from each
of the three sets associated to ``\emph{pease}'', ``\emph{porridge}'' and ``\emph{hot} OR \emph{cold}'',
and keep the minimal intervals among those spanned by such triples (see
Figure~\ref{fig:example}):
\begin{multline*}
\{\?[0\..2],[1\..3],[2\..4],[3\..5],[4\..6],[5\..7],[6\..17],[7\..31],\\
[21\..32],[31\..33],[32\..34],[33\..35],[34\..36]\?\}.
\end{multline*}
From this rich semantic information, a number of different outputs can be
computed. A simple snippet extraction algorithm would
compute greedily the first $k$ smallest nonoverlapping intervals of the antichain, which would yield, for
$k=3$, the intervals $[0\..2]$, $[3\..5]$, $[31\..33]$, that is, ``\emph{Pease}
\emph{porridge} \emph{hot}!'', ``\emph{Pease} \emph{porridge} \emph{cold}!'',
and, again,  ``\emph{Pease} \emph{porridge} \emph{hot}!''.

A ranking scheme such as that proposed by Clarke and Cormack~\cite{ClCSSRR}
would use the number and the length of these intervals to assign a score to the document with
respect to the query. In a simplified setting, we can
assume that each interval yields a score that is the inverse of its length. The
resulting score for the query above would be
\begin{multline*}
\frac1{|[0\..2]|}+\frac1{|[1\..3]|}+\cdots+\frac1{|[6\..17]|}+\frac1{|[7\..31]|}+\cdots
\frac1{|[34\..36]|} \\
=\frac13+\frac13+\cdots+\frac1{12}+\frac1{25}+\cdots\frac13=\frac{177}{50}=3.54.
\end{multline*}
Clearly, documents with a large number of intervals are more relevant, and
short intervals increase the document score more than long intervals. The score
associated to ``\emph{hot}'' would be just $3$ (i.e., the number of
occurrences). One can also incorporate positional information, making, for
example, intervals appearing earlier in the document more
important~\cite{BoVTREC2005}.

What happens if we are asked for a word or a phrase \emph{not} appearing in a
document? In this case, the natural semantics for the query turns out to
be the top element of the lattice, which contains a single empty witness: this
is indeed intuitively appropriate for the semantics of a query that is \emph{not} true in the document---the only
witness is located nowhere. 
This choice is practical, too, as queries of the form $p \text{ AND NOT }
 q$ are true when $p$ is true and $q$ is false, and their witnesses
are the witnesses of $p$, as the top is the unit of conjunction.
More generally, negation should send
all non-bottom elements to bottom, and bottom to top~\cite{BoVLAFI2}, with the
idea that every non-bottom element (every nonempty antichain) represents
(a different degree of) the Boolean value ``true'', 
whereas the bottom is the only representation of ``false''. From an algorithmic viewpoint, implementing
negation is trivial and will not be discussed further.

\section{Operators}
\label{sec:operators}

For the rest of the paper, we assume that we are operating on antichains based
on an unknown total order $O$ for which we just have a comparison operator. We use
$\pm\infty$ to denote a special element that is strictly smaller/larger than all elements
in $O$. Before getting to the core of the paper, however, we highlight the
connection with query resolution in a search engine.

Search engines use inverted lists to index their document
collections~\cite{ZoMIFTSE}. The algorithms described in this paper assume that,
besides the documents in which a term appears, the index makes available the 
\emph{positions} of all occurrences of a term \emph{in increasing order}
(this is a standard assumption).

Given a query, we first obtain the list of documents that could possibly
satisfy the query; this is a routine process that involves merging and intersecting lists.
Once we know that a certain document might satisfy the query, we want to find
its witnesses, if any. To do so, we interpret the terms appearing in the query
as lists of singleton intervals (the term positions), and apply in turn each
operator appearing in the query. The resulting antichain represents the
minimal-interval semantics (i.e., the set of witnesses) of the query with respect to the
document.

For completeness, we define explicitly the operators\footnote{The reader might
be slightly confused by the fact that we are using $\land$ and $\AND$ to denote
essentially the same thing (similarly for $\lor$ and $\OR$). The difference is
that $\land$ is a binary operator, whereas $\AND$ has variable arity. Even if
the evaluation of $\AND$ could be reduced, by associativity, to a composition of
$\land$s, from the viewpoint of the computational effort things are quite
different.} $\AND$ and $\OR$, which
are applied to a list of input antichains \lst Am, resulting in the $\land$ and
$\lor$, respectively, of the antichains \lst Am.
Besides, we consider other useful operators that can be defined directly on the
antichain representation~\cite{CCBASTSFI}. With this aim, let us introduce
a relation $\wellbefore$ between intervals: $I\wellbefore J$ iff $x< y$ for all $x \in I$
and $y \in J$.
 
\begin{enumerate}
\item \emph{(``disjunction operator'')} $\OR$, given input antichains
\lst Am, returns the set of minimal
intervals among those in $A_0\cup A_1\cup\dots\cup A_{m-1}$.
\item \emph{(``conjunction operator'')} $\AND$, given input antichains
\lst Am, returns the set of minimal
intervals among those spanned by the tuples in $A_0\times
A_1\times\cdots\times A_{m-1}$.
\item \emph{(``phrasal operator'')} $\BLOCK$, given input antichains \lst Am,
returns the set of
\emph{intervals} of the form $I_0\cup I_1\cup\cdots\cup I_{m-1}$ with $I_i\in
A_i$ ($0\leq i < m$) and $I_{i-1}\wellbefore I_{i}$ ($0<i<m$).
\item \emph{(``ordered non-overlapping conjunction operator'')} $\ONAND$, given
input antichains \lst Am, returns the set of minimal
intervals among those spanned by the tuples
$\langle I_0,I_1,\ldots,I_{m-1}\rangle \in A_0\times
A_1\times\cdots\times A_{m-1}$ satisfying
$I_{i-1}\wellbefore I_i$.
\item \emph{(``low-pass operator'')} $\LOWPASS_k$,
given an input antichain $A$, returns
the set of intervals from $A$ not longer than $k$.
\item \emph{(``Brouwerian difference\footnote{This operator, denoted by the
minus sign, satisfies the property that $A-B \leq C$ iff $A \leq B \lor C$; it
is sometimes called \emph{pseudo-difference}~\cite{BirLT}.} operator'')} Given two antichains
$A$ (the
minuend) and $B$ (the subtrahend), the difference $A-B$ is the set of intervals
$I \in A$ for which there is no $J \in B$ such that $J \subseteq I$. This
operator was called ``not containing'' in~\cite{CCBASTSFI}.
\item \emph{(Additional containment operators)} Three more operators can be
defined in the same spirit of Brouwerian difference: in~\cite{CCBASTSFI} they
were called ``containing'', ``contained in'' and ``not contained in''. They are
defined, for a pair of antichains $A$ and $B$, as the set of 
intervals of $A$ that, respectively,
\begin{itemize}
  \item contain an interval of $B$;
  \item are contained in an interval of $B$;
  \item are not contained in any interval of $B$. 
\end{itemize}
\end{enumerate}

More informally, given input antichains \lst Am, the operator $\BLOCK$ builds
sequences of \emph{consecutive} intervals, each of which is taken from a
different antichain, in the given order. It can be used, for instance, to implement a
phrase operator. The $\ONAND$ operator is an ordered-$\AND$ operator
that returns intervals spanned by intervals coming from the $A_i$, much like the
$\AND$ operator. However, in the case of $\ONAND$ the left extremes
of the intervals must be nondecreasing, and the
intervals must be nonoverlapping. This operator can be
used, for instance, to search for terms that must appear in a specified order.
$\LOWPASS_k$ restricts the result to intervals shorter than a given
threshold, and be easily combined with $\AND$ or $\ONAND$ to implement searches
for terms that must not be too far apart, and possibly appear in a given order.
Finally, the Brouwerian difference considers the interval in the subtrahend as
``poison'' and returns only those intervals in the minuend that are not
poisoned by any interval in the subtrahend; this operator finds useful
applications, for example, in the case of passage search if the poisoning
intervals are taken to be small (possibly singleton) intervals around the
passage separators (e.g., end-of-paragraph, end-of-sentence, etc.). The
remaining containment operators have similar applications
(see~\cite{CCBASTSFI}).\footnote{
We remark that in a lattice is sometimes possible
to define a relative pseudo-complement operator. This
operator, denoted by an arrow, is the dual of pseudo-difference, and it
satisfies the property that $A\land B \leq C$ iff $A \leq B \to C$~\cite{BirLT}.
However, on one hand this operator has no interpretation in
information-retrieval terms; and, on the other hand, it is easy to show that
$A\to B$ can be an infinite antichain even if $A$ and $B$ are
finite~\cite{BoVLAFI2}.
For these reasons, the computation of relative pseudo-complements will not be pursued in this paper.}

Note that the natural lattice operators $\AND$ and $\OR$ cannot return the
empty antichain when all their inputs are nonempty. This is not true of other
operators: for instance, $\BLOCK$ might fail to find a sequence of
consecutive intervals even if all its inputs are nonempty.

Finally, we remark that all intervals satisfying the definition of the $\BLOCK$
operator are minimal. Indeed, suppose by contradiction that for two
concatenations of minimal intervals we have $[\ell\..r]\subset[\ell'\..r']$
(which implies either $\ell'<\ell$ or
$r<r'$).
Assume that $\ell'<\ell$ (the case $r<r'$ is similar), and note that removing
the first component interval
from both concatenations we still get intervals strictly containing one
another. 
We iterate the process, obtaining two intervals of
$A_{m-1}$ strictly containing one another.

\section{Lazy evaluation}

The main point of this paper is that algorithms for computing operators on
antichain of intervals should be always \emph{lazy} and \emph{linear in the
input intervals}: if an algorithm is lazy, when only a small number of intervals
is needed (e.g., for presenting snippets) the computational cost is
significantly reduced. Moreover, lazy algorithms can be combined in a
hierarchical way to compute lazily the result of any query. 

Linearity in the input intervals is the best possible result for a lazy
algorithm, as input must be read at some point. All algorithms described in this
paper satisfy this property, albeit in the case of $\AND$ and $\OR$ there is
also a logarithmic factor in the number of input antichains.

Note that if the inverted index provides random-access lists of term positions,
algorithms such as those proposed in~\cite{ClCSSRR} might be more appropriate
for first-level operators (e.g., logical operators computed directly on lists of
term positions), as by accessing directly the term positions they achieve
complexity proportional to $ms\log n$, where $n$ is the overall number of term
positions, $m$ is the number of terms, and $s$ is the number of results.
Nonetheless, as soon as one combines several operators, the advantage of an
efficient lazy implementation is again evident, in particular for
automatically expanded queries, in which $m$ can be large.

As we already remarked, in our algorithms we restrict the operations on
the elements of the underlying order $O$ to comparisons. In particular, intervals can be handled
just by comparing their extremes. In this model, the logarithmic factor in the
number of antichains can be easily proved to be unavoidable for the $\OR$
operator: 
\begin{theorem}
Every algorithm to compute $\OR$ that is only allowed to compare interval
extremes requires $\Omega(n \log n)$ 
comparisons for $n$ input intervals.
\end{theorem}
\begin{proof}
It is possible to sort $n$
distinct integers by computing the $\OR$ of $n$ antichains, each made by
just one singleton interval containing one of the integers to be sorted. The
resulting antichain is exactly the list of sorted integers. By an
application of the $\Omega(n\log n)$ lower bound for sorting in this model, we
get the result.
\end{proof}

\subsection{Minimal and optimal laziness}

The term ``lazy'' is usually quoted informally, in particular in the context of functional
or declarative programming. In this paper we consider algorithms that access input antichains
under the form of lists that return the corresponding intervals in their natural order.
We want to define formally a notion of laziness that makes it possible
to prove optimality results. 

We restrict ourselves to algorithms that read
their inputs from an array of lists. Each list is accessible via a  ``next''
function that returns the next element from the list, and when a list is
empty it returns $\NULL$. Analogously, each algorithm has a ``next''
function that returns the next output element (i.e., random access is not
allowed), and when the output is over it returns $\NULL$. So such algorithms can be thought of as producing an output
list that can then be fed to another operator.

Given an algorithm $\A$, an input $I$ (i.e., an array of lists), let us write $\rho^{\A}_i(I,p)$
for the number of elements (including possibly $\NULL$) read by $\A$ from the
$i$-th list of the input array
$I$ when the $p$-th output is produced (sometimes, we will omit $\A$, $I$ or $p$
when they are clear from the context); when writing $\rho^{\A}_i(I,p)$ we
shall always assume that the $0\leq i<m$ (where $m$ is the number of
input
lists) and that the output of $\A$ on input $I$ contains at least $p$ intervals.

\smallskip
\begin{definition}
Two algorithms are \emph{functionally equivalent} iff they produce the
same output list when they are given the same input lists. 
\end{definition}
A first property that we would like our algorithms to feature is that there is
no algorithm that uses strictly less inputs:
\begin{definition}
\label{def:mlazy}
An algorithm $\A$ is \emph{minimally lazy} if,
for every functionally equivalent algorithm $\mathscr B$
such that 
\begin{equation}
\label{eq:minimally1}
\rho^\B_i(I,p)\leq \rho^{\A}_i(I,p)
\end{equation}
for all input $I$, and all $i$ and $p$, we
actually have 
\begin{equation}
\label{eq:minimally2}
\rho^\B_i(I,p) = \rho^{\A}_i(I,p).
\end{equation}
\end{definition}
The property above is very natural, but at the same time it is very weak:
the key point is that~(\ref{eq:minimally2}) must be true \emph{only
of algorithms satisfying}~(\ref{eq:minimally1}). Minimal laziness does not
rule out the existence of an algorithm~$\mathscr B$ that reads one input
element more than $\A$ on a single input, but uses much less input elements on all other
inputs. Nonetheless, minimal laziness embodies the notion that $\A$ cannot be
improved ``locally'', that is, it cannot be improved for some input without
making it worse on some other input. 

All algorithms described in this paper will be minimally lazy. However, for 
most of them we will be able to prove an additional property:
\begin{definition}
\label{def:lazy}
An algorithm $\A$ is \emph{$k$-lazy} iff for every functionally equivalent 
algorithm $\mathscr B$, and for all input $I$, and all $i$ and $p$,
we have 
\begin{equation}
\label{eq:opt}
\rho^{\A}_i(I,p)\leq \rho^\B_i(I,p)+k.
\end{equation}
An algorithm $\A$ is \emph{optimally $k$-lazy} if it is $k$-lazy
and there exists no functionally equivalent $(k-1)$-lazy
algorithm. We say it is \emph{optimally lazy} if it is optimally $k$-lazy for
some $k$.
\end{definition}
With respect to Definition~\ref{def:mlazy}, the essential difference
is that~(\ref{eq:opt}) must be true for \emph{all} functionally equivalent
algorithms $\B$.  
Algorithms that are $k$-lazy have some ``looseness'' in their
usage of the input (the parameter $k$), but given that looseness, they beat
every other algorithm. We have to introduce $k$ because some algorithms contain
choices that makes $0$-laziness unattainable (i.e., depending on the
order of the lists in the input array the algorithm will read more from an
input list rather than from another).

If there is a
$k$-lazy algorithm for a problem, there must be a minimum $\bar k$ for which
there is such an algorithm, and $\bar k$-lazy algorithms will be optimally lazy.
Optimally lazy algorithms cannot be improved ``globally'', that is, at the same time for all inputs:
\begin{theorem}
Let $\A$ be an optimally lazy algorithm. Then, there is no functionally
equivalent algorithm $\B$ such that
\begin{equation}
\rho^\B_i(I,p)< \rho^{\A}_i(I,p)
\end{equation}
for all input $I$, and all $i$ and $p$.
\end{theorem}
\begin{proof}
By contradiction, suppose $\A$ is optimally $k$-lazy for some $k$ and $\B$ is as
in the statement. Then, for every algorithm $\mathscr C$
\[
\rho^{\B}_i(I,p)<\rho^{\A}_i(I,p)\leq \rho^{\mathscr C}_i(I,p)+k,
\]
which implies that $\B$ is $(k-1)$-lazy, contradicting the optimality of $\A$.
\end{proof}
Note that the converse is not true: an algorithm that cannot be improved
globally might not be optimally lazy. However, since by definition
there are no $k$-lazy algorithms when $k$ is negative, $0$-lazy 
algorithms are optimally lazy. Moreover, $0$-laziness
implies trivially minimal laziness. It is also easy to see that for no
$k$ minimal laziness implies (optimal) $k$-laziness.
Our aim is at algorithms that are minimally \emph{and} optimally lazy.

Another way of interpreting the notion of ``being optimally lazy'' is the following: 
let us say that $\A$ undergoes a loss of $k$ on the triple $\langle I,i,p\rangle$ if there is another
functionally equivalent algorithm $\B$ that reads $k$ input elements less
for the same triple. The global loss of $\A$ is the supremum of the losses on all triples $\langle I,i,p\rangle$:
$\A$ is $k$-lazy if its global loss is $k$ (or less). An algorithm is optimally lazy iff it has
the smallest possible (finite) global loss.

There is a subtlety in Definitions~\ref{def:mlazy} and~\ref{def:lazy} that is
worth remarking. By requiring that the parameter $p$ is never greater than the
number of intervals in the output, we are not considering how many elements are
read from the input lists to emit the final $\NULL$. In principle, this choice
implies that even minimally lazy algorithms may consume useless input
elements to emit their final $\NULL$. A more thorough analysis would be required
to include also this case, but it would yield a further subdivision of the
above taxonomy of optimality: indeed, for some problems we consider it is easy
to
show there is no $\NULL$-optimal solution. We think that such an analysis would
add
little value to the present work, as behaving lazily on non-$\NULL$ outputs is a
sufficiently strong property by itself.

\section{General remarks}

In the description and in the proofs of our algorithms, we use interchangeably
$A_i$ to denote the $i$-th input \emph{antichain} and the \emph{list}
returning its intervals in their natural order (and, ultimately, $\NULL$). This
ambiguity should cause no difficulty to the reader.

To simplify the exposition, in the pseudocode we often test whether a list is
empty. Of course, this is not allowed by our model, but in all such cases the
following instruction retrieves the next interval from the same list. Thus, the
test can be replaced by a call that retrieves the next interval and tests for
$\NULL$.

In all our algorithms, we do not consider the case of inputs equal to the top of
the lattice (the antichain formed by the empty interval). For all our operators, the top
either determines entirely the output (e.g., $\OR$) or it is irrelevant (e.g.,
$\AND$). Analogously, we do not consider the case of inputs equal to the bottom
of the lattice (the empty antichain), which can be handled by a test on the
first input read. 

More generally, when proving optimal laziness, it is common to meet situations
in which an initial check is necessary to rule out obvious outputs.
The initial check can make the algorithm analysis more complicated, as its logic
could be wildly different from the true algorithm behaviour. To simplify this
kind of analysis, we prove the following metatheorem, which covers the cases
just described; in the statement of the theorem, $\A$ represent the algorithm
performing the initial check, whereas $\B$ does the real job:
\begin{theorem}
\label{teo:meta}
Let $\B$ be an algorithm defined on a set of inputs $B$, and $\A$ be defined on
a larger set of inputs $A \supseteq B$, and such that
\begin{itemize} 
      \item on all inputs $I \in B$, $\A$ outputs a one-element list containing a
    special element, say $\bot$, and
      \item for all $I\in B$ and all $i$, $\rho_i^{\A}(I,1)\leq \rho_i^{\B}(I,1)$.
\end{itemize} 
\item Then, there exists an algorithm, denoted by $\A\star\B$, such that
\begin{itemize}
  \item $\A\star\B$ is functionally equivalent to $\B$ on $B$;
  \item $\A\star\B$ is functionally equivalent to $\A$ on $A \setminus B$;
  \item if $\A$ and $\B$ are (minimally) optimally lazy on $A \setminus B$ and
  $B$, respectively, then $\A\star\B$ is (minimally) optimally lazy on $A$.
\end{itemize}
\end{theorem}
\begin{proof}
Algorithm $\A\star\B$ simulates algorithm $\A$ and caches the input read so far. If
$\A$ emits any element different from $\bot$, the simulation goes on until $\A$
is done, without caching the input any longer; otherwise, $\A\star\B$ starts
executing $\B$ on the
cached input and possibly on the remaining part of the input until $\B$ is done.

It is immediate to check that $\A\star\B$ is indeed functionally equivalent to $\A$
and $\B$ on $A \setminus B$ and $B$, respectively, and moreover 
\[
	\rho_i^{\A\star\B}(I,p)=\begin{cases}
                     \rho_i^{\A}(I,p) & \text{if $I \in A \setminus B$}\\
                     \rho_i^{\B}(I,p) & \text{if $I \in B$.}\\
                     \end{cases}
\]
Suppose now that $\A$ is $a$-lazy and $\B$ is $b$-lazy for some minimal $a$ and
$b$, and let $c=\max\{\?a,b\?\}$. For every algorithm $\C$ that is functionally
equivalent to $\A\star\B$, we have that $\rho_i^{\C}(I,p)\leq \rho_i^{\B}(I,p)+b$ for all
$I \in B$, and $\rho_i^{\C}(I,p)\leq \rho_i^{\A}(I,p)+a$ for all $I \in A \setminus
B$. But then, using the observation above, $\rho_i^{\C}(I,p)\leq \rho_i^{\A\star\B}(I,p)+c$ for
all $I \in A$, so $\A\star\B$ is $c$-lazy.

Suppose now that $\C$ is functionally equivalent to $\A\star\B$ but that it is
$(c-1)$-lazy, and assume that $c=b$ (the other case is analogous). Then, for all
$I \in B$, $\rho_i^{\C}(I,p)\leq \rho_i^{\A\star\B}(I,p)+c-1=\rho_i^{\B}(I,p)+b-1$; but since
$\C$ is also functionally equivalent to $\B$ on $B$, the latter inequality
contradicts the minimality of $b$.

For minimal laziness, suppose that $\C$ is functionally equivalent to $\A\star\B$
and such that $\rho_i^{\C}(I,p)\leq \rho_i^{\A\star\B}(I,p)$ for all $I \in A$. In
particular, this means that $\rho_i^{\C}(I,p)\leq \rho_i^{\A}(I,p)$ for all $I \in A
\setminus B$, and $\rho_i^{\C}(I,p)\leq \rho_i^{\B}(I,p)$ for all $I \in B$. The
minimal laziness of $\A$ and $\B$ imply that $\rho_i^{\C}(I,p)=\rho_i^{\A}(I,p)$ for
all $I \in A \setminus B$ and $\rho_i^{\C}(I,p)=\rho_i^{\B}(I,p)$ for all $I \in B$,
hence $\rho_i^{\C}(I,p)=\rho_i^{\A\star\B}(I,p)$ for all $I \in A$.
\end{proof}

Incidentally, we observe that $\A\star\B$ requires in general more space than $\A$
or $\B$, because of caching; nonetheless, in all our applications we will need
to cache just one item per input list.

\section{Algorithms based on queues}
\label{sec:algdipq}

The algorithms we provide for $\AND$ and $\OR$ are inspired by the
plane-sweeping technique used in~\cite{SaIFAkWPS} for their proximity algorithm,
which is on its own right a variant of the standard sorted-list merge. The
algorithms are implemented using a min-priority queue~\cite{KnuACPSS}.

At each time, the queue contains a set of indices representing input lists from
which at least one input has been read, and from which $\NULL$ has not been
read yet. Initially, the queue is empty, and $i$ can be added to the
queue calling the function enqueue($Q$,$i$). Priorities are represented by
intervals. The priority of a list is given by
the last interval read from it: for each algorithm, we will specify a different
order between priorities.

The function dequeue($Q$) removes and returns the list of minimum
priority, whereas top($Q$) returns the minimum priority, that
is, the last interval read from a list of minimum priority; we refer to this
interval as ``the top interval''.
Table~\ref{table:dipq} summarises the operations available on a priority queue.

\begin{table}
\begin{center}
\small
\begin{tabular}{|l|l|}
\hline
enqueue($Q$,$i$) & insert item with index $i$ in the queue\\
top($Q$) & returns the minimum priority\\
dequeue($Q$) & returns the index of an item of minimum priority\\
&and removes it from the queue\\
size($Q$) & returns the number of items currently in the queue\\
\hline
\end{tabular}
\end{center}
\caption{\label{table:dipq}The operations available for a priority queue.}
\end{table}

A trivial array-based implementation requires linear space (in the number of
input lists) and has constant cost for all operations modifying the queue,
whereas retrieving the top requires linear time. A better implementation uses a
heap with linear space, logarithmic time
complexity for all operations modifying the queue and constant-time top
retrieval. 

When using heaps, all algorithms based on priority queues have time complexity
$O(n\log m)$ if the input is formed by $m$ antichains containing $n$ intervals overall,
and use $O(m)$ space. This is immediate, as all loops contain exactly one queue
advancement. The worst-case complexity of an array-based implementation is
instead $O(nm)$.
One should consider carefully which implementation to use, however,
as in the case of a very small arity (e.g., three input lists) the array-based implementation turns out to be
significantly faster in practice.

\subsection{Basic comparators}

Our algorithms will be based on two priority orders. The first one,
denoted by $\ebsuff$, is defined by
\[
[\ell\..r]\ebsuff [\ell'\..r'] \iff \text{$r<r'$ or $(r=r'$ and
$\ell\geq \ell')$.} 
\]
In other words, $[\ell\..r]\ebsuff [\ell'\..r']$ if $[\ell\..r]$ ends before or is
a suffix of $[\ell'\..r']$. Note in particular that (somewhat
counterintuitively) $[\ell\..r]\ebsuff [\ell'\..r]$ iff $\ell\geq \ell'$. 

The second order, denoted by $\sbprol$, is defined by
\[
[\ell\..r]\sbprol [\ell'\..r'] \iff \text{$\ell<\ell'$ or $(\ell=\ell'$ and
$r\geq r')$.} 
\]
In other words, $[\ell\..r]\sbprol [\ell'\..r']$ if $[\ell\..r]$ starts before
or prolongs $[\ell'\..r']$. Note in particular that $[\ell\..r]\sbprol [\ell\..r']$ iff $r\geq r'$, and that the
following implication holds:
\[
[\ell\..r]\subseteq [\ell'\..r'] \implies [\ell\..r]\ebsuff [\ell'\..r'] \text{
and } [\ell'\..r']\sbprol [\ell\..r] \]

The algorithms for $\AND$/$\OR$ use a priority queue with
priority order $\sbprol$ or $\ebsuff$. In the
initialisation phase, we read an interval from each list, and
the queue contains all lists.

To simplify the description, we define a
procedure advance($Q$) that updates
with the next interval a list of minimum priority.
If the update cannot be
performed because the list is empty, the list is dequeued.
The function is described in pseudocode in Algorithm~\ref{alg:advance}.

\begin{Algorithm}
\begin{tabbing}
\setcounter{prgline}{0}
\hspace{0.3cm} \= \hspace{0.3cm} \= \hspace{0.3cm} \= \hspace{0.3cm} \=
\hspace{0.3cm} \=\kill\\
\pl\>\PROCEDURE advance($Q$) \BEGIN\\
\pl\>\>\var{i} $\leftarrow$ dequeue($Q$);\\
\pl\>\>\IF $A_i$ is not empty \THEN\\
\pl\>\>\>next($A_i$);\\
\pl\>\>\>enqueue($Q$,$i$)\\
\pl\>\>\END;\\
\pl\>\END;
\end{tabbing}
\caption{\label{alg:advance}The advance function.}
\end{Algorithm}

\subsection{The $\OR$ operator}
\label{sec:OR}

We start with the simplest nontrivial operator. To compute the $\OR$ of the
antichains \lst Am, we merge them using a priority queue $Q$ with
priority order $\ebsuff$.

We keep track of the last interval $c$ returned (initially, $c=[-\infty\..-\infty]$).
When we want to compute the next interval, we advance $Q$ as long as
the top interval contains $c$, and then if the queue is not empty we return the
top interval.
The algorithm\footnote{Note that this algorithm, as discussed in Section~\ref{sec:previous},
can be derived from the dominance algorithms presented in~\cite{KLPFMSV}.} is
described in pseudocode in Algorithm~\ref{alg:or}.

\begin{theorem}
Algorithm~\ref{alg:or} for $\OR$ is correct.
\end{theorem}
\begin{proof}
First of all, note that all intervals in \lst Am are assigned to \var{c} at some
point, unless they contain a previously returned interval. Thus, we just have to
prove that only minimal intervals are returned.

Let $[\ell\..r]$ be a non-minimal element of $A_0\cup A_1\cup\dots\cup A_{m-1}$, and
$[\ell'\..r']$ the
\emph{largest} (according to $\ebsuff$) minimal interval
contained in $[\ell\..r]$. After returning $[\ell'\..r']$ (which certainly appears at
the top of the queue \emph{before} $[\ell\..r]$ due to the fact that $\subseteq$
implies $\ebsuff$), all intervals in the
queue have a right extreme larger than or equal to $r'$. When we advance the queue,
and until we get past $[\ell\..r]$, the top interval will always contain $[\ell'\..r']$,
for otherwise there would be a minimal interval with right extreme between $r'$ and $r$,
and $[\ell'\..r']$ would not be largest. Thus, the while loop will eventually remove $[\ell\..r]$.

To prove that all returned intervals are unique, we just have to note that when
$I$ is returned, each other copy of $I$ is the last interval read from
some list. Thus, at the next call the while loop will be
repeated until all remaining copies are discarded.
\end{proof}

\begin{Algorithm}
\begin{tabbing}
\setcounter{prgline}{0}
\hspace{0.3cm} \= \hspace{0.3cm} \= \hspace{0.3cm} \= \hspace{0.3cm} \=
\hspace{0.3cm} \=\kill\\
\pl\>\INITIALLY\ \var{c} $\leftarrow [-\infty\..-\infty]$ and $Q$ contains one
interval from each $A_i$.\\
\pl\>\FUNCTION next \BEGIN\\
\pl\>\>\WHILE $Q$ is not empty \LAND\ \var{c} $\subseteq$ top($Q$)\;\DO\\
\pl\>\>\> advance$(Q)$\\
\pl\>\>\END;\\
\pl\>\>\IF $Q$ is empty \THEN \RETURN \NULL;\\
\pl\>\>\var{c} $\leftarrow$ top($Q$);\\
\pl\>\>\RETURN\ \var{c}\\
\pl\>\END;
\end{tabbing}
\caption{\label{alg:or}The algorithm for the $\OR$ operator. Note that the
second part of the while condition is actually equivalent to
``left(top($Q$)) $\leq$ left(\var{c})'' due to the monotonicity of
the top-interval right extreme.}
\end{Algorithm}

\begin{theorem}
Algorithm~\ref{alg:or} for $\OR$ is $0$-lazy (and thus optimally and minimally
lazy).
\end{theorem}
\begin{proof}
The first output of the algorithm (let us call it $\A$) requires reading exactly
one interval from each list. No correct algorithm can emit the first output without this data.

Suppose now that for an algorithm $\A^*$ it happens that
\[
\rho^{\A^*}_i(I,p) < \rho^\A_i(I,p)
\]
for some input $I$ and some $i$ and $p$. Upon returning the $p$-th output
$[\ell\..r]$ we have just read from each list the least
interval (w.r.t.~$\ebsuff$) after $[\ell\..r]$; hence, $\A^*$ emits
$[\ell\..r]$ having read from the $i$-th input list an interval
$[\ell'\..r']$ strictly smaller than $[\ell\..r]$ according to $\ebsuff$; this means that
either $r'<r$, or $r'=r$
and $\ell<\ell'$, but the latter case is ruled out by minimality of $[\ell\..r]$.
Thus, $r'<r$, and $\A^*$ would return an incorrect output if the $i$-th input
list would return $[s\..s]$ as next input, with $r'<s<r$. 
\end{proof}

\subsection{The $\AND$ operator}

The $\AND$ operator is more challenging. The priority order of $Q$ is
$\sbprol$, and additionally the queue keeps track of the largest right
extreme of any interval ever read, which we will
call \emph{the right extreme of $Q$} (we just need a variable that is maximised with the right
extreme of each new input interval). We say that $Q$ is \emph{full} if it contains exactly $m$ indices, where
again $m$ is the number of input antichains.

At any time, the interval \emph{spanned} by $Q$ is the interval defined by the
left extreme of the top interval and the right extreme of $Q$: it
will be denoted by span($Q$). Clearly, it is the minimum interval containing all
intervals currently in the queue.

We keep track of the last interval $c$ returned (initially, $c=[-\infty\..-\infty]$).
When we want to compute the next interval, we first advance $Q$ until the spanned
interval does not contain $c$, and in case $Q$ is no longer full we return
\NULL. Then, we store the interval $[\ell\..r]$ currently spanned by $Q$ as a
candidate and advance $Q$. If the new interval spanned by $Q$ is included in
$[\ell\..r]$ we repeat the operation, updating the candidate. Otherwise (or if $Q$
is no longer full) we just return the candidate. The algorithm is described in
pseudocode in Algorithm~\ref{alg:and}.

\begin{Algorithm}
\begin{tabbing}
\setcounter{prgline}{0}
\hspace{0.5cm} \= \hspace{0.3cm} \= \hspace{0.3cm} \= \hspace{0.3cm} \=
\hspace{0.3cm} \=\kill\\
\pl\>\INITIALLY\ \var{c} $\leftarrow [-\infty\..-\infty]$ and $Q$ contains one
interval from every $A_i$.\\
\pl\>\FUNCTION next \BEGIN\\
\pl\>\>\WHILE $Q$ is full \LAND\ \var{c} $\subseteq$ span($Q$) \DO\\
\pl\>\>\>advance$(Q)$\\
\pl\>\>\END;\\
\pl\>\>\IF $Q$ is not full \THEN \RETURN \NULL;\\
\pl\>\>\DO\\
\pl\>\>\>\var{c} $\leftarrow$ span$(Q)$;\\
\setcounter{ifline}{\theprgline}\pl\>\>\>\IF\ \var{c} $=$ top($Q$) \THEN \RETURN
\ \var{c}\,; \\
\pl\>\>\>advance$(Q)$\\
\setcounter{whileendline}{\theprgline}\pl\>\>\WHILE $Q$ is full \LAND span($Q$) $\subseteq$ \var{c}\,; \\
\pl\>\>\RETURN\ \var{c}\\
\pl\>\END;\\
\end{tabbing}
\caption{\label{alg:and}The algorithm for the $\AND$ operator. Note that 
the second part of the
first while condition can be substituted with ``left(top($Q$)) =
left(\var{c})'' because of the monotonicity of the largest right extreme, and that
the second part of the second while condition can be substituted with
``right(\var{c}) $=$ right($Q$)'' by monotonicity of the top-interval left extreme.}
\end{Algorithm}

\begin{theorem}
\label{teo:and}
Algorithm~\ref{alg:and} for $\AND$ is correct.
\end{theorem}
\begin{proof}
We say that a queue
configuration is \emph{complete} if it contains all copies of the top
interval from all lists that contain it. Now observe that \emph{every
complete configuration of a priority queue is
entirely defined by its top interval}. More precisely, if the top is
an interval $I$ from list $i$, then for every other list $j$ the corresponding
interval $J$ in the queue is the minimum interval in $A_j$ larger than or equal
to $I$ (according to $\sbprol$). Indeed, suppose by contradiction that there is
another interval $K$ from $A_j$ satisfying
\[I\sbprol K\sbprolneq J.\]

Then, at some point $K$ must have entered the queue, and when it has been dequeued
the top must have become some interval $I'\sbprol I$, so we get \[K\sbprol I'\sbprol
I\sbprol K,\] which yields $K=I$: a contradiction, as we assumed the configuration of the
queue to be complete.

We now show that for every minimal interval $[\ell\..r]$ in the $\AND$ of \lst
Am there is a complete configuration of $Q$ spanning $[\ell\..r]$. Consider for
each $i$
the set $C_i$ of intervals of $A_i$ contained in $[\ell\..r]$. At least one of
these sets must contain a (necessarily unique) \emph{right delimiter}, that is, an interval of the
form $[\ell'\..r]$ (see Figure~\ref{fig:and}). Moreover, at least one of the sets containing a
delimiter
must be a singleton. Indeed, if every $C_i$ containing a right delimiter would
also contain some other interval, the right extreme of that interval would
clearly be smaller than $r$: the maximum of such right extremes, say $r'<r$,
would define a spanned interval $[\ell\..r']$ showing that
$[\ell\..r]$ was not minimal. We conclude that at least one $C_i$, say
$C_{\bar\imath}$, is a singleton
containing a right delimiter.

\begin{figure}
\begin{center}
\includegraphics[scale=.7]{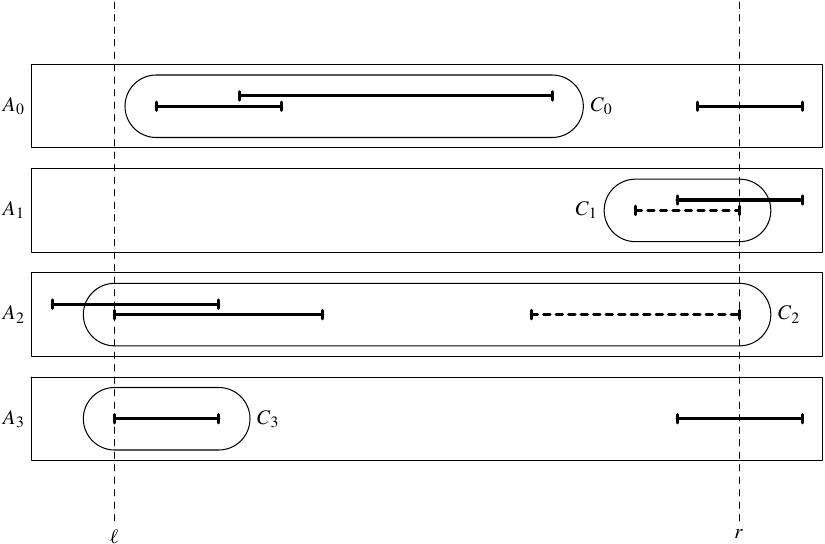}
\end{center}
\caption{\label{fig:and}A sample configuration found in the proofs of
Theorems~\ref{teo:and} and \ref{teo:optand}. The dashed intervals are right
delimiters. The first two input lists are in the inner set; the last two input
lists are in the conflict set; the last input list is also in the resolution set.}
\end{figure}

Let $I_i$ be the leftmost interval in each $C_i$; these intervals are a complete configuration of $Q$: 
if $I_i=[\ell\..r']$ is the $\sbprol$-smallest among such intervals and if $I_i\in A_j$
necessarily $I_i=I_j$, because $A_j$ cannot contain two intervals with the same left extreme.
The set of intervals also spans $[\ell\..r]$ (because the right extreme of $I_{\bar\imath}$ is $r$, and
the left extreme of the $\sbprol$-least interval $I_i$ is $\ell$). We conclude
that all minimal intervals in the output are eventually spanned by $Q$.

However, no minimal interval can be spanned during the first while loop,
unless it has been already returned, as all
intervals spanned in that loop contain a previously returned interval (notice
that at the first call the loop is skipped altogether).
Finally, if an interval is spanned in the second while loop and we do not get
out of the loop, the next candidate interval will be smaller or equal. We
conclude that sooner or later all minimal intervals cause an interruption of the
second while loop, and are thus returned.

We are left to prove that if an interval is returned, it is guaranteed
to be minimal.
If we exit the loop using the check on the top interval, the returned
interval is indeed guaranteed to be minimal.
Otherwise, assume that the interval $[\ell\..r]$ spanned by $Q$ at the start of the
second while loop is not minimal, so $[\bar \ell\..r]\subset[\ell\..r]$, for
some minimal interval $[\bar \ell\..r]$ that will be spanned
later (as we already proved that all minimal intervals
are returned). Since the right extreme of $Q$ is nondecreasing, the second
while loop will pass through intervals of the form $[\ell'\..r]$, with $\ell<\ell'<\bar\ell$, until
we exit the loop.

Finally, we remark the uniqueness of all returned intervals is guaranteed
by the first while loop.
\end{proof}

Note that our algorithm for $\AND$ \emph{cannot} be $0$-lazy,
because the choices made by the queue for equal intervals cause different
behaviours. For instance, on the input lists  
$\{\?[0\..0],[2\..2]\?\}$, $\{\?[1\..1]\?\}$, $\{\?[0\..0],[2\..2]\?\}$
the algorithm advances the last
list before returning $[0\..1]$, but there is a variant of the same algorithm that
keeps intervals sorted lexicographically by $\sbprol$ and by input list index, 
and this variant would advance the first list instead.

Nonetheless:
\begin{theorem}
\label{teo:optand}
Algorithm~\ref{alg:and} for $\AND$ is minimally lazy and optimally $1$-lazy.
\end{theorem}

\begin{proof}
We denote  Algorithm~\ref{alg:and} with $\A$, and let $\A^*$ be a
functionally equivalent algorithm. 
Let us number the intervals appearing in a certain input $I=\text{\lst Am}$: in
particular, let $\bigl[\ell_i^j\..r_i^j\bigr]$ be the $j$-th interval
appearing in $A_i$. For sake of simplicity, let us identify the
$\NULL$ returned as last element by the input lists with the interval
$[\infty\..\infty]$ (it is immediate to see that $\A$
behaves identically). Let us write $\rho_i$ (respectively, $\rho^*_i$) for
$\rho_i^{\A}(I,p)$ (respectively, $\rho_i^{\A^*}(I,p)$), and $[\ell\..r]$
be the $p$-th output interval; let
also $s_i$ be the index of the first interval in list $A_i$ that is
included in $[\ell\..r]$.

We divide the indices of the input lists in two sets: the \emph{inner set}
is the set of indices $i$ for which $\ell<\ell_i^{s_i}$ (that is, the first
interval of  $A_i$ included in $[\ell\..r]$ has left extreme larger than
$\ell$); the \emph{conflict set} is the set of indices $i$ for which
$\ell=\ell_i^{s_i}$ (that is, the first interval of $A_i$ included in
$[\ell\..r]$ has left extreme equal to $\ell$). Finally, the \emph{resolution set}
is a subset of the conflict set containing those indices $i$ for which
$r_i^{s_i+1}>r$ (that is, the successor of the first interval of $A_i$ included in
$[\ell\..r]$ is no longer contained in $[\ell\..r]$). Note that the
resolution set is always nonempty, or otherwise $[\ell\..r]$ would not be
minimal (recall that we substituted $\NULL$ with $[\infty\..\infty]$). The
situation is depicted in Figure~\ref{fig:and}.

We remark the following facts:
\begin{renumerate}
 \item\label{lab:general} for all $i$, $\rho^*_i\geq s_i$; that is, when $\A^*$ outputs $[\ell\..r]$
it has read at least the first interval of the antichain with left extreme larger than or equal
to $\ell$; otherwise, $\A^*$ would emit $[\ell\..r]$ even on a modified
input in which $A_i$ has no intervals contained in $[\ell\..r]$ (such
intervals have index equal to or greater than $s_i$, so they have not been
seen by $\A^*$, yet);
 \item\label{lab:inner} for all $i$ in the inner set, $\rho_i=s_i\leq \rho^*_i$;
 \item\label{lab:conflict} for all $i$ in the conflict set, $\rho_i \in \{s_i,s_i+1\}$; that is,
 in the case an antichain \emph{does} contain an interval $J$ with left extreme
 $\ell$, either the last interval read by $\A$ when $[\ell\..r]$ is output is
 exactly $J$, or it is the interval just after $J$;
 \item\label{lab:top} 
 if for some $i$ we have $\bigl[\ell_i^{s_i}\..r_i^{s_i} \bigr]=[\ell\..r]$, then
$\rho_j=s_j$ for all $j$, because we exit the second while loop at line \theifline;
\item\label{lab:resolution} otherwise,
there is a unique index $\bar\imath$
in the resolution set such that $\rho_{\bar\imath}=s_{\bar\imath}+1$ (i.e.,
$r_{\bar\imath}^{\rho_{\bar\imath}}>r$), and for all other resolution indices $i$ we have
$\rho_i=s_i$ (i.e., $r_i^{\rho_i}\leq r$); this happens because we interrupt the
second while loop when we see the
first interval whose right extreme exceeds $r$ (at line \thewhileendline).
\end{renumerate}

Let us first prove that $\A$ is $1$-lazy by showing that $\rho_i\leq
\rho_i^*+1$: this is true for all indices in the inner set because of~\ref{lab:inner}, 
and for all indices in the conflict set because of $\rho_i\leq
s_i+1\leq \rho_i^*+1$ (by~\ref{lab:conflict} and~\ref{lab:general}).

Now, let us show that $\A^*$ cannot be $0$-lazy. Suppose it is
such; then, in particular, $\rho_i^*\leq \rho_i$ for all indices $i$, and 
we can assume w.l.o.g.~that $\rho_i^* < \rho_i$ for some $i$ (if for all inputs, all
output prefixes and all $i$ we had $\rho_i^*=\rho_i$, then we would conclude that $\A$
is $0$-lazy as well, contradicting the observation made before this theorem).

Note that we can also assume w.l.o.g.~not to be in case~\ref{lab:top}
(as in that case $\rho_i=\rho^*_i$ for all $i$), which also implies that $\ell\neq r$.
Thus, the unique index $\bar\imath$ of~\ref{lab:resolution} is
also the only index
in the resolution set such that $\rho^*_{\bar\imath}=s_{\bar\imath}+1$ ($\A^*$ must
advance some list in the resolution set, or it would emit a wrong output on a modified input
in which the $(s_i+1)$-th interval of $A_i$ is $[r\..r]$ for all $i$ in the conflict set).

Let \lst it be the indices in the conflict set
for which $\rho_{i_p}=s_{i_p}+1$, in the order in which they are accessed from the
corresponding lists by $\A$: clearly $i_{t-1}=\bar\imath$ is the only
resolution index in this sequence, by \ref{lab:resolution}. Let \lst ju be the
indices in the
conflict set for which $\rho^*_{j_p}=s_{j_p}+1$, in the order in which they
are accessed from the corresponding lists by $\A^*$. Necessarily, 
$\{\?j_0,j_1,\dots,j_{u-1}\?\}\subseteq\{\?i_0,i_1,\dots,i_{t-1}\?\}$ (because
$s_{j_p}+1=\rho_{j_p}^*\leq \rho_{j_p} \leq s_{j_p}+1$, hence $\rho_{j_p} = s_{j_p}+1$) 
and inclusion is strict (because, for some index $i$, $\rho_i^*<\rho_i$, hence
$s_i\leq \rho_i^*<\rho_i\leq s_i+1$, which
implies that $i=i_v$ for some $v$, whereas $i\neq j_v$ for all $v$). 

Let $p$ be the
first position that $\A$
and $\A^*$ choose differently, that is, $i_p\neq j_p$ (this happens at least
at the position of \lst ju where $\bar\imath$ appears).
We build a new input similar to \lst Am, except for
$A_{i_p}$ and $A_{j_p}$, which are identical
up to their interval of left extreme $\ell$; then, $A_{i_p}$
continues with $[r'\..r']$ for some $r'>r$ (so $i_p$ is in the resolution set),
whereas
$A_{j_p}$ continues with $[r\..r]$ (so $j_p$ is in the inner set). On this
input, to output $[\ell\..r]$
$\A$ advances the input list $A_{j_p}$ strictly less than $\A^*$, which contradicts
the assumption on $\A^*$.
\end{proof}

\section{Greedy algorithms}

The remaining operators admit greedy algorithms: they advance the input lists
until some condition becomes true. The case
of $\LOWPASS_k$ is of course trivial, and the algorithm for $\BLOCK$ is essentially a
restatement in terms of intervals of the folklore algorithm for phrasal queries.
They are minimally and optimally lazy. The case of $\ONAND$ and Brouwerian
difference are more interesting: $\ONAND$ is the only
algorithm for which we prove the impossibility of an optimally lazy
implementation in the general case.

All greedy algorithms have time complexity $O(n)$ if the input is formed
by $m$ antichains containing $n$ intervals overall, and use $O(m)$ space. This
is immediate, as all loops advance at least one input list.

\subsection{The $\BLOCK$ operator}

The $\BLOCK$ operator is the only one that can be implemented exclusively if the
underlying total order is locally finite,\footnote{A partially ordered set is
\emph{locally finite} if all intervals of the form $[x\..y]$ are finite.} that
is, if it admits a notion of successor.
In discussing this algorithm, we shall assume that every element $x \in O$ has a
successor, denoted by $x+1$, satisfying $x<x+1$ and $x\leq y\leq x+1 \implies
x=y \text{ or } y=x+1$.
 
We keep track of a current interval for all lists \lst Am; initially, these
intervals are set to $[-\infty\..-\infty]$.
When we want to compute the next interval, we update the interval associated to
the first list. Then, we try to fix index $i$ (initially, $i=1$). To do so, we
advance the list $A_i$ until the returned interval has left extreme larger
than the right extreme of the current interval for $A_{i-1}$. If we go too far,
we just advance the first list, reset $i$ to $1$ and restart the process,
otherwise we increment $i$. When we find an interval for $A_{m-1}$ we return the
interval spanned by all current intervals. The algorithm is described in
pseudocode in Algorithm~\ref{alg:block}.

\begin{Algorithm}
\begin{tabbing}
\setcounter{prgline}{0}
\hspace{0.5cm} \= \hspace{0.3cm} \= \hspace{0.3cm} \= \hspace{0.3cm} \=
\hspace{0.3cm} \= \hspace{0.3cm} \=\kill\\
\pl\>\INITIALLY $[\ell_k\..r_k]\leftarrow [-\infty\..-\infty]$ for all $0\leq k< m$.\\ \pl\>\FUNCTION next \BEGIN\\
\pl\>\>\IF $A_0$ is empty \THEN \RETURN \NULL;\\
\pl\>\>$[\ell_0\..r_0]$ $\leftarrow$ next($A_0$);\\
\pl\>\>$i$ $\leftarrow$ 1;\\
\setcounter{externalwhileblock}{\theprgline}\pl\>\>\WHILE $i<m$ \DO\\
\setcounter{internalwhileblock}{\theprgline}\pl\>\>\>\WHILE  $\ell_i\leq r_{i-1}$ \DO \\
\pl\>\>\>\>\IF $A_i$ is empty \THEN \RETURN \NULL;\\
\pl\>\>\>\>$[\ell_i\..r_i]$ $\leftarrow$ next($A_i$)\\
\pl\>\>\>\END;\\
\pl\>\>\>\IF $\ell_i=r_{i-1}+1$ \THEN $i\leftarrow i+1$\\
\pl\>\>\>\ELSE \BEGIN\\
\pl\>\>\>\>\IF $A_0$ is empty \THEN \RETURN \NULL;\\
\pl\>\>\>\>$[\ell_0\..r_0]$ $\leftarrow$ next($A_0$);\\
\pl\>\>\>\> $i\leftarrow 1$\\
\pl\>\>\>\END \\
\pl\>\>\END;\\
\pl\>\>\RETURN $[\ell_0\..r_{m-1}]$\\
\pl\>\END;
\end{tabbing}
\caption{\label{alg:block}The algorithm for the $\BLOCK$ operator.}
\end{Algorithm}

\begin{theorem}
Algorithm~\ref{alg:block} for $\BLOCK$ is correct.
\end{theorem}
\begin{proof}
At the start of an iteration of the external while loop (line \theexternalwhileblock) with a certain
index $i$
we clearly have $r_k+1=\ell_{k+1}$ for $k=0,1,\dots,i-2$. Thus, if we complete
the execution of the loop we certainly return a correct interval.

To complete the proof, we start by proving the following invariant property: at
line \theexternalwhileblock, for all $0<j<m$ there are no intervals in
$A_j$ with left extreme in $[r_{j-1}+1\..\ell_j-1]$. In other words, the $j$-th
current interval $[\ell_j\..r_j]$ has either left extreme smaller than or equal
to $r_{j-1}$, or it is the first interval in $A_j$ whose left extreme is larger
than $r_{j-1}$. The property is trivially true at the beginning, and advancing
$[\ell_0\..r_0]$ cannot change this fact. We are left to prove that the
execution of the internal while loop (line \theinternalwhileblock) cannot either.

During the execution of the loop at line \theinternalwhileblock, only $[\ell_i\..r_i]$ can change.
This affects the invariant because it modifies the intervals
$[r_{i-1}+1\..\ell_i-1]$ and $[r_i+1\..\ell_{i+1}-1]$, but in the second case
the interval is made smaller, so the invariant is \textit{a fortiori} true. In
the first case, at the beginning of the execution of the internal while loop
either $r_{i-1}+1\leq \ell_i-1$, that is, $r_{i-1}< \ell_i$, so the loop is not
executed at all and the invariant cannot change, or $r_{i-1}+1>\ell_i-1$, which
means that the interval $[r_{i-1}+1\..\ell_i-1]$ is empty, and the loop will
advance $[\ell_i\..r_i]$ up to the first interval in $A_i$ with a left extreme
larger than $r_{i-1}$, making again the invariant true.

Suppose now that there are $[\bar\ell_0\..\bar r_0]$,~$[\bar\ell_1\..\bar
r_1]$, $\dots\,,$~$[\bar\ell_k\..\bar r_k]$ satisfying $\bar r_i+1=\bar \ell_{i+1}$ for
some $k>0$ and $0\leq i<k$. We prove by induction
on $k$ that at some point during the execution of the algorithm we will be at
the start of the external while loop with $i=k$ and
$[\ell_j\..r_j]=[\bar\ell_j\..\bar r_j]$ for $j=0,1,\dots,k$. The thesis is
trivially true for $k=0$. Assume the thesis for $k-1$, so we are at the start of
the external while loop with $i=k-1$ and $\ell_j=\bar \ell_j$, $r_j=\bar r_j$ for
$j=0,1,\dots k-1$. Because of the invariant, either $[\ell_k\..r_k]=[\bar
\ell_k\..\bar r_k]$ or $[\ell_k\..r_k]$ will be advanced by the execution of the
internal while loop up to $[\bar \ell_k\..\bar r_k]$. Thus, at the end of the
external while loop the thesis will be true for $k$. We conclude that all
concatenations of intervals from \lst Am are returned.

We note that all intervals returned are unique (minimality has been already
discussed in Section~\ref{sec:operators}), as $[\ell_0\..r_0]$ is advanced at
each call, so a duplicate returned interval would imply the existence of two
comparable intervals in $A_0$.
\end{proof}

\begin{theorem}
Algorithm~\ref{alg:block} for $\BLOCK$ is $0$-lazy (and thus optimally and minimally
lazy).
\end{theorem}
\begin{proof}
The algorithm is trivially minimally lazy, as all outputs are uniquely
determined by a tuple of intervals from the inputs. An algorithm $\A^*$ advancing an input
list $A_i$ less than Algorithm~\ref{alg:block}
for some output $[\ell\..r]$ would emit $[\ell\..r]$ even if we truncated $A_i$ after the last
interval read by $\A^*$.
\end{proof}

\paragraph{Practical remarks.} In the case of intervals of integers, the
advancement of the first list at the end of the outer loop can actually be
iterated until $r_0\geq\ell_i-i$. This change does not affect the complexity
of the algorithm, but it may reduce the number of iterations of the outer loop.
In case
the input antichains are entirely formed by singletons\footnote{We emphasise this
case because this is what happens with phrasal queries all of whose
subqueries are simple terms; implementation may treat this special case
differently to obtain further optimisation, for instance using \textit{ad hoc} 
indices~\cite{WZBFPQCI}.}, a folklore algorithm
aligns the singletons circularly rather than starting from the first one (since
they are singletons, once the position of an interval is fixed all the remaining ones
are, too). The main advantage is that of avoiding to resolve several alignments
if the first few terms appear often consecutively, but not followed by the
remaining ones. 

\subsection{The $\ONAND$ operator}

The algorithm for computing this operator is a medley of the algorithms for
$\AND$ and for $\BLOCK$: as in the case of $\AND$, we must check that future
intervals are not smaller then our current candidate $[\ell'\..r']$; as in the case of
$\BLOCK$, there is no queue and the lists \lst Am are advanced greedily.
Again, we keep track of a current interval $[\ell_i\..r_i]$ for every list $A_i$; 
initially, these intervals are $[-\infty\..-\infty]$, except for the first
one, which is taken from the first list.
The algorithm is described in pseudocode in Algorithm~\ref{alg:oand}; an
informal description follows. 

\begin{Algorithm}
\begin{tabbing}
\setcounter{prgline}{0}
\hspace{0.5cm} \= \hspace{0.3cm} \= \hspace{0.3cm} \= \hspace{0.3cm} \=
\hspace{0.3cm} \= \hspace{0.3cm} \=\kill\\
\pl\>\INITIALLY $[\ell_0\..r_0]$ $\leftarrow$ next($A_0$), $[\ell_k\..r_k]\leftarrow
[-\infty\..-\infty]$ for all $0<k<m$ and $i\leftarrow 1$.\\
\pl\>\FUNCTION next
\BEGIN\\ 
\pl\>\>$[\ell'\..r']$ $\leftarrow$ $[\infty\..\infty]$;\\
\pl\>\>\var{b} $\leftarrow\infty$;\\
\pl\>\>\FOREVER\\
\setcounter{whileexit}{\theprgline}\pl\>\>\>\FOREVER\\
\pl\>\>\>\>\IF $r_{i-1}\geq b$ \THEN \RETURN $[\ell'\..r']$;\\
\setcounter{exitouterloop}{\theprgline}\pl\>\>\>\>\IF $i=m$ \LOR $\ell_i> r_{i-1}$ \THEN \BREAK;\\
\setcounter{innerwhileloop}{\theprgline}\pl\>\>\>\> \DO \\ 
\pl\>\>\>\>\>\IF $r_i \geq \text{\var{b}}$ \LOR $A_i$ is empty \THEN \RETURN
$[\ell'\..r']$;\\
\pl\>\>\>\>\>$[\ell_i\..r_i]\leftarrow$ next($A_i$)\\ \pl\>\>\>\>\WHILE $\ell_i\leq r_{i-1}$;\\
\pl\>\>\>\>$i\leftarrow i+1$;\\
\pl\>\>\>\END;\\
\setcounter{middlewhileend}{\theprgline}\pl\>\>\>$[\ell'\..r']$ $\leftarrow$ $[\ell_0\..r_{m-1}]$;\\
\pl\>\>\>$b\leftarrow \ell_{m-1}$;\\
\pl\>\>\>$i\leftarrow 1$;\\
\pl\>\>\>\IF $A_0$ is empty \THEN \RETURN $[\ell'\..r']$;\\
\pl\>\>\>$[\ell_0\..r_0]$ $\leftarrow$ next($A_0$)\\
\pl\>\>\END;\\
\pl\>\END;
\end{tabbing}
\caption{\label{alg:oand}The algorithm for the $\ONAND$ operator. For sake of
simplicity, we
use the convention that returning $[\infty\..\infty]$ means returning \NULL, and
that if one of the input lists is exhausted the function returns \NULL.}
\end{Algorithm}

The core of the algorithm is in the loop starting at
line~\theinnerwhileloop: this loop tries to align the $i$-th interval, that is,
advance it until $[\ell_{i-1}\..r_{i-1}]\wellbefore [\ell_i\..r_i]$. The loop starting at
line~\thewhileexit\ aims at aligning all
intervals; note that we assume as an invariant that, after the first execution,
every time we discover that the $i$-th interval is already aligned we can
conclude that also the remaining intervals (the ones with index larger than $i$)
are aligned as well (second condition at line~\theexitouterloop).

The loop at line~\thewhileexit\ can be interrupted as soon as, trying to align
the $i$-th interval, we exhaust the $i$-th list or we find an interval whose right
extremes exceeds $b$, the left extreme of the $(m-1)$-th interval forming the current
candidate alignment. If any such condition is satisfied, the current candidate
is certainly minimal and can thus be returned.

Upon a successful alignment (line~\themiddlewhileend), we have a new candidate:
note that either this is the first candidate (i.e.,
$[\ell'\..r']=[\infty\..\infty]$ before the assignment), or its right extreme coincides
with the one of the previous candidate (i.e., $r'=r_{m-1}$ before the assignment), whereas
its left extreme is certainly strictly larger.
In either case, we try to see if we can advance the first interval and find a
new, smaller candidate with a new alignment: this should explain the outer loop.

\begin{theorem}
\label{teo:oand}
Algorithm~\ref{alg:oand} for $\ONAND$ is correct.
\end{theorem}
\begin{proof}
Let us say that a sequence
$[\ell'_h\..r'_h]\wellbefore [\ell'_{h+1}\..r'_{h+1}]\wellbefore \cdots
\wellbefore[\ell'_{k-1}\..r'_{k-1}]$ of intervals ($h<k\leq m$), one from
each list $A_h,A_{h+1},\ldots,A_{k-1}$, is \emph{leftmost} if, for all $h<j<k$, there are
no intervals in $A_j$ with left extreme in $(r'_{j-1}\..\ell'_j)$: such a
sequence is uniquely determined by $k$ and by $[\ell'_h\..r'_h]$.
Let $[\bar\ell\..\bar r]$ be the interval returned at the last call (initially,
$[\bar\ell\..\bar r]=[\infty\..\infty]$). Then, the following invariant holds
at the start of the loop at line~\thewhileexit:
\begin{enumerate}
  \item  $[\ell_0\..r_0]\wellbefore [\ell_1\..r_1]\wellbefore \cdots
\wellbefore[\ell_{i-1}\..r_{i-1}]$ is leftmost;
\item if $\ell_i\neq-\infty$ also
$[\ell_i\..r_i]\wellbefore[\ell_{i+1}\..r_{i+1}]\wellbefore\cdots\wellbefore[\ell_{m-1}\..r_{m-1}]$ is
leftmost;
\item if $[\ell_{i-1}\..r_{i-1}]\wellbefore[\ell_{i}\..r_{i}]$ then this pair is
leftmost.
\end{enumerate}
The fact
that this invariant holds is easy to check; in particular, see the inner while
loop at line~\theinnerwhileloop\ and the exit at line~\theexitouterloop.

We now show that each output interval $[\bar\ell\..\bar r]$ is at some time
assigned to $[\ell'\..r']$. Note that $i>0$ at all times, so $[\ell_0\..r_0]$ is
assigned only at the end of the infinite loop. This means that $[\ell_0\..r_0]$
runs through the whole first input list. 

Thus, as soon as $\ell_0=\bar\ell$ the inner loop will either compute the
leftmost representation of $[\bar\ell\..\bar r]$, or exit prematurely. In the
second case, the function will necessarily complete the leftmost representation at the
next call. We conclude that leftmost representations of all output intervals are
assigned to $[\ell'\..r']$ eventually: since $[\bar\ell\..\bar r]$ is minimal,
it will be emitted before $[\ell'\..r']$ is assigned again. Uniqueness follows
by uniqueness of leftmost representations.
\end{proof}

It is not difficult to see that there is \emph{no algorithm} for $\ONAND$ that is
$k$-lazy for any $k$, except for the case $m=2$; indeed:

\begin{theorem}
\label{teo:nooand}
If $m>2$, there exist no optimally lazy algorithm for $\ONAND$.
\end{theorem}
\begin{proof}
By contradiction, let $\B$ be $k$-lazy, and observe that, on any
given input $I$, every algorithm for
$\ONAND$, before emitting its $p$-th output $[\ell\..r]$, must have reached at least
the leftmost sequence $[\ell'_0\..r'_0]\wellbefore [\ell'_1\..r'_1]\wellbefore
\cdots \wellbefore[\ell'_{m-1}\..r'_{m-1}]$ spanning it. Now, choose any 
$x \in (r'_{m-2}\..\ell'_{m-1})$ and, for all
$i=0,\dots,m-2$, take an arbitrary sequence $u_i^0<u_i^1<u_i^2<\dots<u_i^{k+1}
\in (r'_i\..\min\{\?\ell'_{i+1},x\?\})$; also choose an arbitrary sequence
$v^0<v^1<v^2<\dots<v^{k+1} \in (x\..\ell'_{m-1})$. Run $\B$ on a different input
$J$, obtained as follows: whenever $\B$ asks for an input from list $i<m-1$, 
we use the original intervals from $I$ only up to $[\ell'_i\..r'_i]$, and then
we do the following: if $i<m-1$, we start offering $[u_i^0\..v^0]$,
$[u_i^1\..v^1]$ and so on; as far as
the last input list is concerned, we do not make any change. An example of this
construction is given in Figure~\ref{fig:oand}.

\begin{figure}
\begin{center}
\includegraphics[scale=.7]{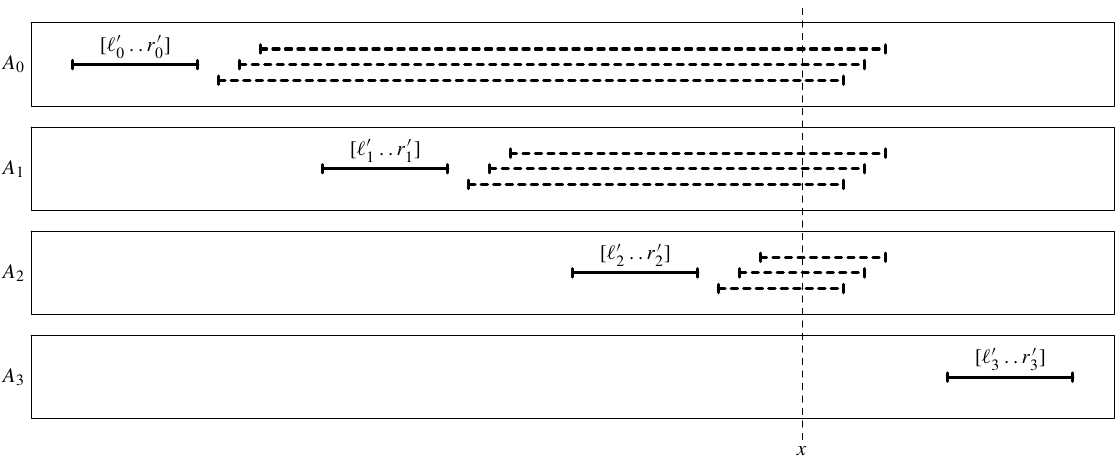}
\end{center}
\caption{\label{fig:oand}A sample configuration found in the proof of
Theorem~\ref{teo:nooand}. In this case, $m=4$ and $k=1$. The dashed intervals
are those of the form $\bigl[u_i^j\..v^j\bigr]$: while reading such intervals it
is impossible to decide whether the continuous intervals span an element of the
output.}
\end{figure}

Note that the intervals are chosen so that $\B$ cannot yet emit $[\ell\..r]$,
because there is always some chance for it not to be minimal. 
We stop testing $\B$ as soon as, for some $\bar\imath$, $\B$ has
read at least $k+2$ inputs after $[\ell'_{\bar\imath}\..r'_{\bar\imath}]$ from 
the $\bar\imath$-th list for some $\bar\imath<m-1$; let $J'$ be the portion of $J$ read by $\B$
so far, let $\bar\jmath$ any index different from $\bar\imath$
and from $m-1$, and let $\A$ be an algorithm for $\ONAND$ obtained from $\B$ by
modifying its behaviour on the input as follows: when faced with an input that
coincides with $I$ up to $[\ell'_0\..r'_0]\wellbefore [\ell'_1\..r'_1]\wellbefore
\cdots \wellbefore[\ell'_{m-1}\..r'_{m-1}]$ inclusive, it then reads one more
interval for each list and, if all these intervals contain any common point, say
$z$, it starts reading from list $\bar\jmath$ until an interval not including
$z$ is reached, or until the $\bar\jmath$-th list ends, in which case it emits $[\ell\..r]$.
Note that this modification does not harm the correctness of the algorithm, but
now $\rho^{\A}_{\bar\imath}(J',p)+k+1=\rho^{\B}_{\bar\imath}(J',p)$ which contradicts
the $k$-laziness of $\B$.
\end{proof}

Hence, for $\ONAND$, there is no hope for our algorithm to be  
optimally lazy in the general case; yet, it enjoys three interesting properties:

\begin{theorem}
\label{teo:quasiopt}
Let $\A$ be Algorithm~\ref{alg:oand} for $\ONAND$.
\begin{enumerate} 
 \item\label{enu:minlaz} $\A$ is minimally lazy;
 \item\label{enu:optlaz} $\A$ is $0$-lazy (and thus optimally and minimally
 lazy) when $m=2$;
 \item\label{enu:twoinputs} for any functionally equivalent algorithm $\B$,
 $\rho^{\A}_i(I,p)\leq \rho^{\B}_i(I,p+1)$; that is, our algorithm, to produce
 any output, never reads more input than $\B$ needs to produce its next output.
\end{enumerate}
\end{theorem}
\begin{proof}
\noindent(\ref{enu:minlaz}) Suppose that $\B$ is functionally equivalent to $\A$
and $\rho_j^{\B}(I,p)\leq \rho_j^{\A}(I,p)$ for every $j$, $I$ and $p$, and
$\rho_{\bar\jmath}^{\B}(\bar I,\bar p)<\rho_{\bar\jmath}^{\A}(\bar I,\bar p)$ for some
specific $\bar\jmath$, $\bar I$ and $\bar p$. Let $[\ell\..r]$ be the $\bar
p$-th output on input $\bar I$, and $[\ell'_0\..r'_0]\wellbefore\dots
\wellbefore [\ell'_{m-1}\..r'_{m-1}]$ be its leftmost spanning sequence
(Figure~\ref{fig:quasiopt} displays an example); when $\A$ outputs
$[\ell\..r]$, we have that $[\ell_j\..r_j]=[\ell'_j\..r'_j]$ for all $j>i$,
whereas the $i$-th list is over or is such that $r_i\geq \ell'_{m-1}$ (with leftmost $r_i$),
$[\ell_0\..r_0]\wellbefore \dots \wellbefore [\ell_{i-1}\..r_{i-1}]$ is
leftmost and $\ell<\ell_0$. Since no correct algorithm can emit $[\ell\..r]$
before scanning its input up to the leftmost spanning sequence, necessarily
$\bar\jmath\leq i$. 

Moreover, necessarily $\bar\jmath\neq i$: otherwise, we could
modify the inputs by substituting the unread intervals of
the lists $A_i,A_{i+1},\ldots,A_{m-2}$ with a suitable sequence of
aligned intervals which, together with the remaining ones, would span
$[\ell_0\..r]$; this would make $[\ell\..r]$ non minimal. 
 
Now,
 suppose that $J$ is an input equal to $\bar I$ but modified so that the
 $\bar\jmath$-th
 list ends immediately after the last interval read by $\B$: on input $J$,
 algorithm $\A$ does not read a single interval from list $i$ beyond 
 $[\ell'_i\..r'_i]$, because it emits $[\ell\..r]$ as soon as the test for
 emptiness of $A_{\bar\jmath}$ is performed. So $\rho_i^{\A}(J,\bar 
 p)<\rho_i^{\B}(J,\bar p)$, a contradiction. 

\begin{figure}
\begin{center}
\includegraphics[scale=.7]{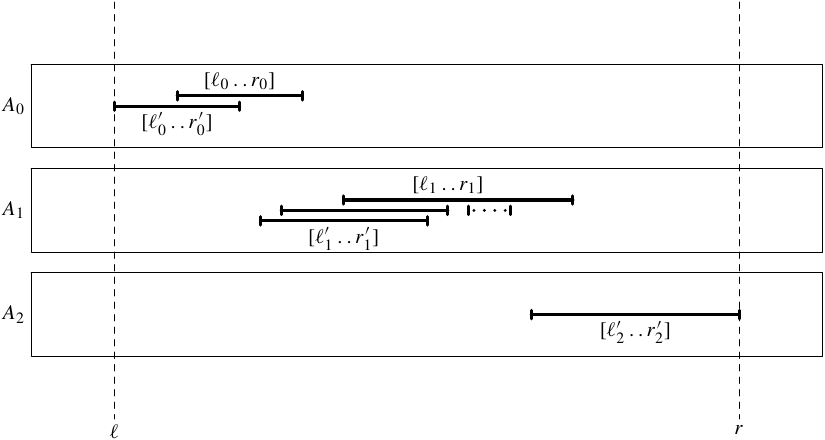}
\end{center}
\caption{\label{fig:quasiopt}A sample configuration found in the proof of
Theorem~\ref{teo:quasiopt}. The intervals $[\ell'_i\..r_i]$ form a leftmost spanning
sequence, and $i=1$, so $\bar\jmath=0$. Note that no algorithm can avoid reading
$[\ell_1\..r_1]$, or it would fail if we replaced it with the dotted interval.}
\end{figure}

\noindent(\ref{enu:optlaz}) We prove that $\A$ is $0$-lazy in that case.
Indeed, when a certain output $[\ell\..r]$ is ready to be produced, $\A$ 
tries to read one more interval $[\ell_0\..r_0]$ from the first list, and this
is unavoidable (any other algorithm must do this, or otherwise we might modify
the next interval so that $[\ell\..r]$ is not minimal). This interval has a
right extreme larger than or equal to $\ell_1$, or otherwise $[\ell\..r]$ would
not be minimal: $\A$ exits at this point, so it is $0$-lazy.

\noindent(\ref{enu:twoinputs}) This is trivial: when $\A$ outputs an interval,
it has not yet reached (or, it has just reached) the leftmost sequence spanning
the following output, and no correct algorithm could ever emit the next output
before that point.
\end{proof}

\paragraph{Practical remarks.} In the case of intervals of integers, the check
for $r_i\geq b$ can replaced by $r_i\geq b-(m-i-2)$, and the check for
$r_{i-1}\geq b$ by $r_i\geq b-(m-i-1)$, obtaining in some case faster detection
of minimality. If the input antichains are entirely formed by singletons, the check $r_i\geq
b$ can be removed altogether, because in that case we know that $r_i=\ell_i\leq
r_{i-1}<b$.

\subsection{Brouwerian difference}

The Brouwerian difference $M-S$ between antichains $M$ (the minuend) and $S$
(the subtrahend) can be computed
by searching greedily, for each interval $[\ell\..r]$ in $M$, the first
interval $[\ell'\..r']$ in $S$ for which $\ell'\geq \ell$ or $r'\geq r$. We keep track of the
last interval $[\ell'\..r']$ read from the input list $S$ (initially,
$[\ell'\..r']=[-\infty\..-\infty]$) and update it until $\ell'\geq \ell$ or $r'\geq r$. At that
point, if we did not exhaust $S$ and $[\ell'\..r']\subseteq[\ell\..r]$ (in which
case $[\ell\..r]$ should not be output) we
continue scanning $M$; otherwise, we return $[\ell\..r]$. The algorithm is described in
pseudocode in Algorithm~\ref{alg:diff}.

\begin{Algorithm}
\begin{tabbing}
\setcounter{prgline}{0}
\hspace{0.5cm} \= \hspace{0.3cm} \= \hspace{0.3cm} \= \hspace{0.3cm} \=
\hspace{0.3cm} \= \hspace{0.3cm} \=\kill\\
\pl\>\INITIALLY $[\ell'\..r']\leftarrow [-\infty\..-\infty]$.\\ 
\pl\>\FUNCTION next \BEGIN\\
\setcounter{brouwerone}{\theprgline}\pl\>\>\WHILE $M$ is not empty \DO\\
\pl\>\>\>$[\ell\..r]$ $\leftarrow$ next($M$);\\
\setcounter{brouwertwo}{\theprgline}\pl\>\>\>\WHILE $S$ is not empty \LAND $\ell'< \ell$ \LAND $r'< r$ \DO\\
\pl\>\>\>\>$[\ell'\..r']$ $\leftarrow$ next($S$)\\
\pl\>\>\>\END; \\
\setcounter{brouwerthree}{\theprgline}\pl\>\>\>\IF $S$ is empty \LOR $[\ell'\..r']\not\subseteq[\ell\..r]$ \THEN \RETURN $[\ell\..r]$\\
\pl\>\>\END;\\
\pl\>\>\RETURN \NULL\\
\pl\>\END;
\end{tabbing}
\caption{\label{alg:diff}The algorithm for Brouwerian difference (a.k.a.~``not containing'').}
\end{Algorithm}

\begin{theorem}
Algorithm~\ref{alg:diff} for Brouwerian difference is correct.
\end{theorem}
\begin{proof}
Note that at the start of the inner while loop (line \thebrouwertwo) 
$[\ell'\..r']$ contains
either the leftmost interval of $S$ such that $\ell'\geq\ell$ or $r'\geq r$,
or some interval preceding it. This is certainly true at the first call, and
remains
true after the execution of the inner while loop because of the first part of
its exit condition (line \thebrouwertwo). Finally, advancing the list of $M$ cannot make the
invariant false.

Given the invariant, at the end of the inner loop $[\ell'\..r']$ contains the
leftmost interval of $S$ such that $\ell'\geq \ell$ or $r'\geq r$, if such
an interval exists. Note that if $[\ell'\..r']$ is not contained in
$[\ell\..r]$, then no other interval of $S$ is. Indeed, if $\ell'<\ell$ this
means that $r'\geq r$, so all preceding intervals have too small left extremes,
and all following intervals have too large right extremes (the same happens
\textit{a fortiori} if $\ell'\geq\ell$).
Thus, the test at line \thebrouwerthree\ will emit $[\ell\..r]$ if and only if it
belongs to the output.
\end{proof}

\begin{theorem}
Algorithm~\ref{alg:diff} for Brouwerian difference is $0$-lazy (and thus
optimally and minimally lazy).
\end{theorem}
\begin{proof}
When Algorithm~\ref{alg:diff} outputs $[\ell\..r]$, it has read just
just $[\ell\..r]$ from $M$ and the first element
$[\ell'\..r']$ of $S$ such that $\ell'\geq \ell$ or $r'\geq r$. If either interval
has not been read by some other algorithm $\A$, $\A$ would fail if we
removed altogether $[\ell\..r]$ from $M$ or if we substituted $[\ell'\..r']$
with $[\ell\..r]$ and deleted all following intervals in $S$.
\end{proof}

\subsection{Other containment operators}

The three remaining containment operators have greedy, minimally lazy algorithms
similar to Algorithm~\ref{alg:diff}, and are shown as
Algorithm~\ref{alg:contains}, \ref{alg:iscontained}
and~\ref{alg:isnotcontained}.
The correctness and $0$-laziness of the algorithms can
be easily derived along the lines of the proofs for Brouwerian difference.

\begin{Algorithm}
\begin{tabbing}
\setcounter{prgline}{0}
\hspace{0.5cm} \= \hspace{0.3cm} \= \hspace{0.3cm} \= \hspace{0.3cm} \=
\hspace{0.3cm} \= \hspace{0.3cm} \=\kill\\
\pl\>\INITIALLY $[\ell'\..r']\leftarrow [-\infty\..-\infty]$.\\ 
\pl\>\FUNCTION next \BEGIN\\
\pl\>\>\WHILE $A$ is not empty \DO\\
\pl\>\>\>$[\ell\..r]$ $\leftarrow$ next($A$);\\
\pl\>\>\>\WHILE $B$ is not empty  \LAND $\ell'< \ell$ \LAND $r'< r$ \DO\\
\pl\>\>\>\>$[\ell'\..r']$ $\leftarrow$ next($B$)\\
\pl\>\>\>\END; \\
\pl\>\>\>\IF $B$ is empty \THEN \RETURN \NULL;\\
\pl\>\>\>\IF$[\ell'\..r']\subseteq[\ell\..r]$ \THEN\RETURN $[\ell\..r]$\\
\pl\>\>\END;\\
\pl\>\>\RETURN \NULL\\
\pl\>\END;
\end{tabbing}
\caption{\label{alg:contains}The algorithm for the ``containing''
operator.}
\end{Algorithm}

\begin{Algorithm}
\begin{tabbing}
\setcounter{prgline}{0}
\hspace{0.5cm} \= \hspace{0.3cm} \= \hspace{0.3cm} \= \hspace{0.3cm} \=
\hspace{0.3cm} \= \hspace{0.3cm} \=\kill\\
\pl\>\INITIALLY $[\ell'\..r']\leftarrow [-\infty\..-\infty]$.\\ 
\pl\>\FUNCTION next \BEGIN\\
\pl\>\>\WHILE $A$ is not empty \DO\\
\pl\>\>\>$[\ell\..r]$ $\leftarrow$ next($A$);\\
\pl\>\>\>\WHILE $B$ is not empty \LAND   $r'< r$ \DO\\
\pl\>\>\>\>$[\ell'\..r']$ $\leftarrow$ next($B$)\\
\pl\>\>\>\END; \\
\pl\>\>\>\IF $B$ is empty \THEN \RETURN \NULL;\\
\pl\>\>\>\IF $\ell' \leq \ell$ \THEN \RETURN $[\ell\..r]$\\
\pl\>\>\END;\\
\pl\>\>\RETURN \NULL\\
\pl\>\END;
\end{tabbing}
\caption{\label{alg:iscontained}The algorithm for the ``is contained''
operator.}
\end{Algorithm}

\begin{Algorithm}
\begin{tabbing}
\setcounter{prgline}{0}
\hspace{0.5cm} \= \hspace{0.3cm} \= \hspace{0.3cm} \= \hspace{0.3cm} \=
\hspace{0.3cm} \= \hspace{0.3cm} \=\kill\\
\pl\>\INITIALLY $[\ell'\..r']\leftarrow [-\infty\..-\infty]$.\\ 
\pl\>\FUNCTION next \BEGIN\\
\pl\>\>\WHILE $A$ is not empty \DO\\
\pl\>\>\>$[\ell\..r]$ $\leftarrow$ next($A$);\\
\pl\>\>\>\WHILE $B$ is not empty \LAND $r'< r$  \DO\\
\pl\>\>\>\>$[\ell'\..r']$ $\leftarrow$ next($B$)\\
\pl\>\>\>\END; \\
\pl\>\>\>\IF $B$ is empty \LOR $\ell < \ell'$ \THEN \RETURN$[\ell\..r]$\\
\pl\>\>\END;\\
\pl\>\>\RETURN \NULL\\
\pl\>\END;
\end{tabbing}
\caption{\label{alg:isnotcontained}The algorithm for the ``is not contained''
operator.}
\end{Algorithm}

\section{Previous work}
\label{sec:previous}

The only attempt at linear lazy algorithms for
minimal-interval region algebras we are aware of is the work of Young--Lai and
Tompa on \emph{structure selection
queries}~\cite{YoTOERAE}, a special type of expressions built on the primitives
``contained-in'', ``overlaps'', and so on, that can be evaluated lazily in
linear time. Their motivations are similar to ours---application of region
algebras to very large text collections. Similarly, Navarro and
Baeza--Yates~\cite{NaBCLAPSS} propose a class of
algorithms that using tree-traversals are able to compute efficiently several
operations on overlapping regions. Their motivations are efficient
implementation of structured query languages that permit such regions.
Albeit similar in spirit, they do not
provide algorithms for any of the operators we consider, and they do not provide
a formal proof of laziness.

The manipulation of antichain of intervals can be translated into manipulation
of points in the plane compared by \emph{dominance}---coordinatewise
ordering. Indeed, $[\ell\..r]\supseteq[\ell'\..r']$ iff
the point $(\ell,-r)$ is dominated by the point $(\ell',-r')$. Dominance problems have been
studied for a long time in computational geometry: for instance,~\cite{KLPFMSV}
presents an algorithm to compute the maximal elements w.r.t.~dominance. This
method can be turned into an algorithm for antichains of intervals by coupling
it with a simple (right-extreme based) merge to produce an algorithm for the
$\OR$ operator. One has just to notice that since dominance is symmetric in the extremes,
the mapping $[\ell\..r]\mapsto(-r,\ell)$ turns minimal intervals (by containment)
into maximal points (by dominance). The algorithm described in~\cite{KLPFMSV}
assume a decreasing first-coordinate order of the points, which however is an
\emph{increasing} ordering by right extreme on the original intervals. After
some cleanup, the algorithm turns out to be identical to our algorithm for $\OR$
(albeit the authors do not study its laziness).

The other operators have no significant geometric meaning, and to the best of
our knowledge there is no algorithm in computational geometry that computes them.

Lazy evaluation is a by-now classical topic in the theory of computation, dating
back to the mid-70s~\cite{HMLE}, originally introduced for expressing the
semantics of call-by-need in functional languages. However, the notion of lazy
optimality used in this paper is new, and we believe that it captures as
precisely as possible the idea of optimality in accessing sequentially multiple
lists of inputs in a lazy fashion.

\section{Conclusions}

We have provided efficient lazy algorithms for the computation of several
operators on the lattice of interval antichains. The algorithms for lattice operations
require time $O(n\log m)$ for $m$ input antichains containing $n$ intervals
overall, whereas the remaining algorithms are linear in $n$.  In particular, the
algorithm for $\OR$
has been proved to be optimal in a comparison-based model. Moreover, the
algorithms are minimally and optimally lazy (with the exception of $\ONAND$ when
$m>2$, in which case we prove an impossibility result) and use space linear in the
number of input antichains.

We remark that, in principle, input antichains need not be finite. As long as
the underlying order is locally finite and the ``next''
operator returns more intervals that form an antichain (ordered by their extremes), the algorithms described in this paper will return more results. In this sense, they can be thought as algorithms that transform infinite
input streams into infinite output streams.

An interesting open problem is that of providing a matching lower bound for the
$\AND$ operator (in the comparison-based computational model).

\bibliography{biblio}

\end{document}